\documentclass[fleqn,usenatbib]{mnras}

\usepackage[T1]{fontenc}
\usepackage{ae,aecompl}

\usepackage{bm}
\usepackage{graphicx}	
\usepackage{amsmath}	
\usepackage{amssymb}	
\usepackage{adjustbox}
\usepackage{comment}
\usepackage{siunitx}
\usepackage{float}
\usepackage[table,xcdraw]{xcolor}
\usepackage{xspace}
\usepackage{etoolbox}
\usepackage{multirow}
\usepackage{tcolorbox}
\usepackage{soul}

\usepackage{newtxtext,newtxmath}

\usepackage{todonotes}
\defcitealias{Wegg19}{W19}

\newtcolorbox{mybox}{colback=red!5!white,colframe=red!75!black}
\newtcolorbox{boxcontent}[1]{colback=gray!5!white,colframe=gray!75!black,fonttitle=\bfseries,title=#1}

\newcommand{\Msun}{\mbox{M$_{\odot}$}}

\newcommand{\vir}[1]{``#1''}

\newcommand {\dd}{\mathrm d}

\newcommand{\gaia}{\textit{Gaia} }

\definecolor{mygreen}{rgb}{0.0, 0.42, 0.24}

\definecolor{abcolor}{rgb}{0.0, 0.25, 0.5}

\title[RR Lyrae from Binary Evolution]{RR Lyrae From Binary Evolution: Abundant, Young and Metal-Rich}

\author[A.Bobrick et al.]{Alexey Bobrick$^{1}$\thanks{bobrick@campus.technion.ac.il --- equally contributing co-first author},
Giuliano Iorio$^{2,3,4}$\thanks{giuliano.iorio.astro@gmail.com --- equally contributing co-first author},
Vasily Belokurov$^{5,6}$,
Joris Vos$^{7}$,
\newauthor
Maja Vu\v{c}kovi\'{c}$^{8}$,
Nicola Giacobbo$^{9}$
\\
$^{1}$Technion - Israel Institute of Technology, Physics Department, Haifa, Israel 32000\\
$^{2}$Dipartimento di Fisica e Astronomia “Galileo Galilei”, Università di Padova, vicolo dell’Osservatorio 3, IT-35122, Padova, Italy \\
$^{3}$INAF - Osservatorio Astronomico di Padova, vicolo dell’Osservatorio 5, IT-35122 Padova, Italy dell’Osservatorio 3, IT-35122, Padova, Italy \\
$^{4}$INFN - Padova, Via Marzolo 8, I--35131 Padova, Italy\\
$^{5}$Institute of Astronomy, University of Cambridge, Madingley Road, Cambridge CB3 0HA, UK\\
$^6$Center for Computational Astrophysics, Flatiron Institute, 162 5th Avenue, New York, NY 10010, USA\\
$^{7}$Astronomical Institute of the Czech Academy of Sciences, CZ-25165, Ondřejov, Czech Republic\\
$^{8}$Instituto de F\'{\i}sica y Astronom\'{\i}a, Universidad de Valpara\'{\i}so, Gran Breta\~{n}a 1111, Playa Ancha, Valpara\'{\i}so 2360102, Chile\\
$^{9}$School of Physics and Astronomy \& Institute for Gravitational Wave Astronomy, University of Birmingham, Birmingham, B15 2TT, UK\\
}

\date{Accepted XXX. Received YYY; in original form ZZZ}

\pubyear{2023}

\begin{document}
\label{firstpage}
\pagerange{\pageref{firstpage}--\pageref{lastpage}}
\maketitle

\begin{abstract}
RR Lyrae are a well-known class of pulsating horizontal branch stars widely used as tracers of old, metal-poor stellar populations. However, mounting observational evidence shows that a significant fraction of these stars may be young and metal-rich. Here, through detailed binary stellar evolution modelling, we show that all such metal-rich RR Lyrae can be naturally produced through binary interactions. Binary companions of these RR Lyrae stars formed through binary interactions partly strip their progenitor's envelopes during a preceding red giant phase. As a result, stripped horizontal branch stars become bluer than their isolated stellar evolution counterparts and thus end up in the instability strip. In contrast, in the single evolution scenario, the stars can attain such colours only at large age and low metallicity. While binary-made RR Lyrae can possess any ages and metallicities, their Galactic population is relatively young ($1$~--~$9\,{\rm Gyr}$) and dominated by the Thin Disc and the Bulge. We show that Galactic RR Lyrae from binary evolution are produced at rates compatible with the observed metal-rich population and have consistent G-band magnitudes, Galactic kinematics and pulsation properties. Furthermore, these systems dominate the RR Lyrae population in the Solar Neighbourhood. We predict that all metal-rich RR Lyrae have an A, F, G or K-type companion with a long orbital period ($P \gtrsim 1000\,{\rm d}$). Observationally characterising the orbital periods and masses of such stellar companions will provide valuable new constraints on mass and angular momentum-loss efficiency for Sun-like accretors and the nature of RR Lyrae populations.
\end{abstract}




\section{Introduction}
\label{sec:Introduction}

\subsection{Classical RR Lyrae}

RR Lyrae variables are one of the best-known types of stellar pulsators \citep{Catelan2004, Catelan2009}. These helium-burning horizontal branch (HB) stars are believed to come from an old ($>10\,{\rm Gyr}$) metal-poor ($\mathrm{[Fe/H]}\lesssim-1$) stellar population. They are located in the instability strip (IS) in a compact region of the Hertzsprung-Russell (HR) diagram. Their chemistry, evolution and lightcurve properties are well understood, making these stars an excellent probe of stellar pulsations and stellar evolution \citep{Smith2004}. Since RR Lyrae are easy to identify, one can use them to discover faint old metal-poor objects such as dwarf galaxies or globular clusters, e.g. \citet{Sesar2014, Torrealba2019}. 

Thanks to their relatively narrow colour range, RR Lyrae have been used to measure reddening and probe the interstellar medium (ISM), e.g. \citet{Haschke2011}. Similarly, since their luminosity at a given metallicity is accurately known, RR Lyrae are one of the primary standard candles. The accuracy of RR Lyrae as standard candles has recently been improved thanks to the data from {\it Gaia} \citep{Muraveva2018,GaiaVariable}. As standard candles, RR Lyrae stars have been used to measure the distances to globular clusters and nearby galaxies with $\sim$2\% accuracy, e.g. \citet{Sarajedini2006, Braga2015}. They have also been used in constructing the cosmological distance ladder for nearby galaxies, complementing the distances from the brighter Cepheid variables \citep{Beaton2016, Riess2016}. Furthermore, since RR Lyrae belong to an old Galactic stellar population, they have been extensively used to characterise the substructures in the Galactic Halo and reconstruct the Milky Way merger history, e.g. \citet{Fiorentino2015, Belokurov2017, Fiorentino17, Iorio18, deBoer2018, Iorio19}. Finally, RR Lyrae pulsators have provided a wealth of information on the structures and formation histories of the Galactic Thick Disc, Bulge, Magellanic Clouds, dwarf spheroidals and other galaxies in the { Local Group}, e.g. { \citet{Feast2008, Pietrukowicz2015, Gran2015, Ablimit17, Ablimit18, Monelli18, MuravevaReticulum18, MuravevaDraco20, Cusano21, IB21,Ablimit22}}.

The physics behind the RR Lyrae pulsations is well understood \citep{Smith2004}. An HB star consists of a hot helium-burning core surrounded by a relatively thin and cold hydrogen envelope. The envelope reprocesses the radiation from the stellar core, thus shifting its spectrum redwards. Therefore, the mass of the envelope and its opacity dictate the effective temperature of the star and its position along the HB. The RR Lyrae variability, as well as the variability of other stars in the IS, is produced by radial pulsation of the stellar layers driven by the $\kappa-\gamma$ mechanism. In the stellar interior, the H and He layers are partially ionized and act as a “thermodynamic” valve trapping energy during compression and releasing it during expansion, allowing the full development of radial oscillations \citep[][and references therein]{CatelanBook}.  

Low-mass Sun-like stars ($M_{\rm init}\lesssim 2\,\Msun$) go through the He core flash before landing on the HB. Therefore, on the HB, they have similar He-cores ($M_{\rm core}\approx 0.5\,\Msun$) with similar luminosity nearly independent of the progenitor mass, e.g. \citet{Rood73, Castellani81, Caputo1987}. Hence, the temperature and colour of HB stars depend mainly on the amount of H-envelope and its opacity. The opacity, in turn, is sensitive to metallicity, with more metal-rich stars being more opaque, better at reprocessing radiation and hence redder. Therefore, metal-poor stars ($\mathrm{[Fe/H]}\lesssim -1.0$) with initial masses of $0.7$~--~$0.8\,\Msun$ have sufficiently low-mass envelopes and sufficiently low opacity so that they become hot enough to land in the IS. Since the typical main-sequence lifetimes of such stars are long, $>10\,{\rm Gyr}$, these are old population-II stars. Such old metal-poor \vir{classical} RR Lyrae represent the bulk of the RR Lyrae in the Stellar Halo and globular clusters.

\subsection{Metal-Rich RR Lyrae}

\begin{figure*}
\centering
\includegraphics[width=\linewidth]{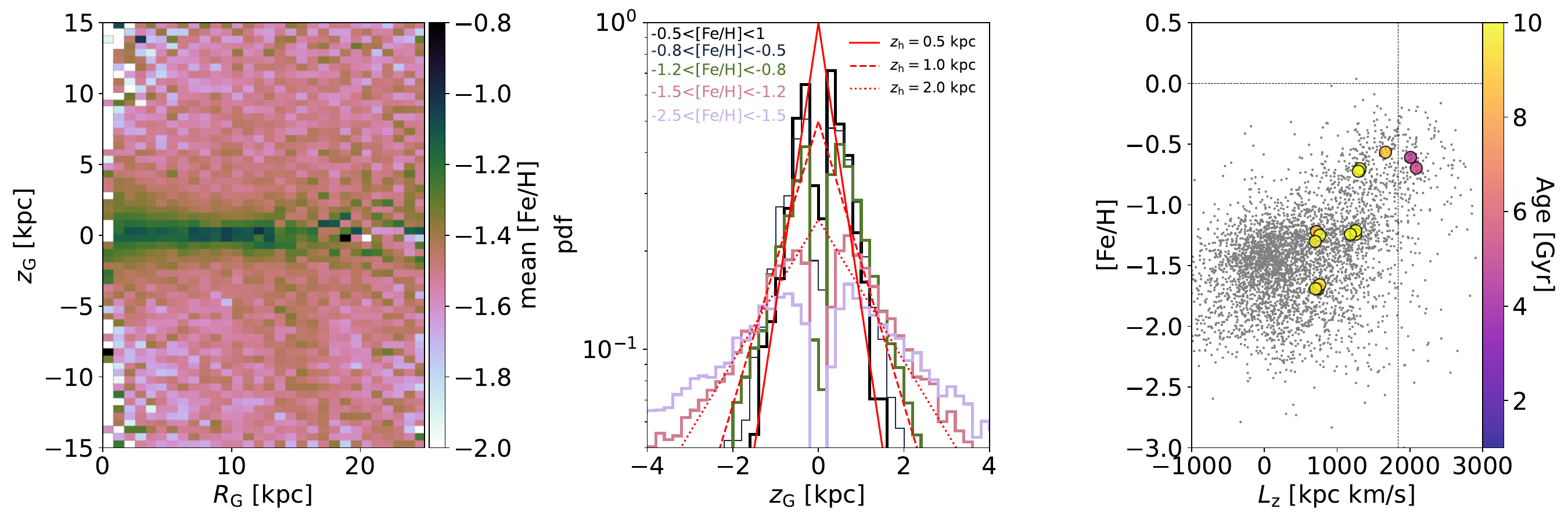}
\caption{Properties of RR Lyrae stars in the {\it Gaia} DR3 catalogue.
{\sl Left panel}: Mean metallicity in bins of  Galactic cylindrical radius $R_\mathrm{G}$  and height above the Galactic plane $z_\mathrm{G}$. Note the presence of a highly flattened and metal-rich population at $z_\mathrm{G}=0$.
{\sl Middle panel}: Vertical distribution of RR Lyrae in different metallicity bins. The red lines show model exponential distributions with scale heights: $0.5\,{\rm kpc}$ (solid); $1.0\,{\rm kpc}$ (dashed); $2.0\,{\rm kpc}$ (dotted).  
{\sl Right panel}: Distribution of {\it Gaia} DR3 RR Lyrae (grey dots) in the $L_z$-$\mathrm{[Fe/H]}$ plane. The coloured circles indicate the population's age derived from the vertical velocity dispersion, based on the \citet{Sharma20} relations. Each circle indicates a given bin in $L_\mathrm{z}$, absolute Galactic height $|z_\mathrm{G}|$ and metallicity, and only bins with more than $30$ stars are shown. The positions of the circles indicate the median value of $L_\mathrm{z}$ and the metallicity of the stars in the bin. The dashed gray lines indicate the solar values for reference, $\mathrm{[Fe/H]}=0$ and $L_\mathrm{z}=1840\,{\rm kpc}\cdot{\rm km}\cdot{\rm s}^{-1}$. 
}
\label{fig:gaiadr3}
\end{figure*}

It is hard to produce metal-rich RR Lyrae within the classical scenario. The atmospheres of old metal-rich HB stars are more opaque to the radiation and appear redder than otherwise similar old, metal-poor stars. Therefore, they must have a less massive envelope to enter the IS. Since more metal-rich stars, according to single stellar evolution models, evolve more slowly\footnote{The stellar cores in metal-rich low-mass stars are larger and colder than in their metal-poor counterparts. Consequently, the efficiency of hydrogen burning is lower, and the metal-rich stars live longer. According to the MIST stellar tracks (\citealt{Mista}, see Section \ref{sec:mod}), a star with initial mass $0.9\,{\rm M}_\odot$ leaves the main sequence approximately after $7\,{\rm Gyr}$ for $\mathrm{[Fe/H]} =-2$, after $13\,{\rm Gyr}$ for $\mathrm{[Fe/H]}=0$, and after $16\,{\rm Gyr}$ for $\mathrm{[Fe/H]}=0.3$. Since interacting stars do not follow single stellar evolution, this reasoning does not apply to binary-made RR Lyrae discussed in Section~\ref{sec:BinMetalRichIntro}}, and need a smaller initial mass to have thinner envelopes on the HB, there is an age-metallicity correlation among classical RR Lyrae, with higher-metallicity classical RR Lyrae being older (see, e.g. \citealt{Savino20}). Metal-rich stars can, at least in principle, enter the IS either by starting with a lower initial mass ($M_{\rm init}\lesssim 0.7\,\Msun$) or via a significant mass loss on the Red Giant Branch (RGB). However, as we discuss in detail in Section~\ref{sec:Discussion}, the absence of non-interacting evolved stars less massive than $0.7\,\Msun$ and the observational constraints on the RGB mass loss prevent RR Lyrae formation at metallicities above $\mathrm{[Fe/H]}\gtrsim-0.5$.

On the other hand, the observations of the Galactic metal-rich RR Lyrae challenge the classical formation scenario. Despite RR Lyrae stars traditionally being interpreted as old, metal-poor Population II stars, the existence of metal-rich RR Lyrae stars in the vicinity of the Sun is well documented. These objects reach metallicities as high as $\mathrm{[Fe/H]}\approx0.2$  and their vertical Galactic distribution, kinematics  \citep{Layden1995a,Layden1995b,Maintz05, Marsakov18,Liu13,Prudil20,Zinn20,Gilligan21}, and chemical abundances are consistent with the Thin Disc \citep{Marsakov2019,Crestani21}, although there are hints of an abundance deficiency for some elements such as aluminium \citep{Feuillet22} and scandium \citep{Gozha21}.

\cite{IB21} used the unprecedented RR Lyrae catalogue from {\it Gaia} DR2, confirming that the population of metal-rich RR Lyrae is not limited to the Solar neighbourhood but permeates the Thin Disc from the inner parts ($\approx 3$ kpc) out to the disc's outskirts ($\approx 25$ kpc). Surprisingly, the kinematics of this population (circular velocity and velocity dispersions) is consistent with the young Thin Disc ($\lesssim 7$ Gyr, \citealt{Ablimit20,Sharma20}, see also \citealt{Prudil20}). The existence of such a metal-rich and likely young population is a conundrum. Either these RR Lyrae are indeed young, challenging the classical formation scenario of RR Lyrae stars, or they are one of the oldest Milky Way populations that somehow remained completely unperturbed during the turbulent past of our Galaxy \citep[see, e.g.][]{Belokurov2018,Myeong2019}. The existence of such a population is also in tension with the recent discovery of an ancient pre-disk state of our Galaxy \citep[see][]{Aurora,Conroy2022,Myeong2022}.

The third {\it Gaia} data release (GDR3, \citealt{GDR3}) further supports the presence of a large population of RR Lyrae in the young Thin Disc. The new data release includes a dedicated catalogue of more than 270 000 RR Lyrae \citep{RRLG3}. About half of this sample contains a complete characterisation of the lightcurves (pulsation periods and Fourier transform parameters) that can be used to retrieve the photometric metallicity and a robust estimate of distance (see, e.g., \citealt{IB21,Mullen21,Li22}). 

In Figure~\ref{fig:gaiadr3}, we show an overview of the chemical, spatial, and kinematic properties of this sample in the left, middle, and right panels, respectively. The lack of sources at $z_\mathrm{G}\approx0$ is an observational bias due to an increase in the Galactic extinction.  In the right panel, each circle represents a bin in $L_z$ with edges at $500$, $1000$, $1500$, $1800$, $2500$, $3000\,{\rm kpc}\cdot{\rm km}\cdot{\rm s}^{-1}$, absolute Galactic height $|z_\mathrm{G}|$, with bin edges at $0$, $0.5$, $1.0$, $1.5\,{\rm kpc}$, and metallicity, with bin edges at $-3.0$, $-2.0$, $-1.5$, $-1.0$, $1.0$. The velocity dispersions have been estimated by minimizing a Gaussian likelihood with centroid $\langle V_\mathrm{z} \rangle =0$ and considering the contribution of the velocity uncertainties. In the first two panels, we use all the $130557$ RR Lyrae classified as RRab or RRc with an available estimate of the pulsation period and other light curve parameters (e.g. $\phi_\mathrm{31}$) and with a reddening value ($ebv$) smaller than $2$ (based on the \citealt{Schlafly11} reddening map). In the third panel, we use a subsample of $4039$ RR Lyrae satisfying the conditions described above and with available radial velocities in the GDR3 catalogue. The metallicities and the distance have been estimated as in \citet{IB21}. The Sun is assumed to be located at $(x_\mathrm{G}, y_\mathrm{G}, z_\mathrm{G}) = (8.13,0,0)\,{\rm kpc}$ in the Galactic Cartesian coordinates \citep{Gravity18,IB21}, with cylindrical velocities  $(V_\mathrm{R_\mathrm{G}}, V_\mathrm{\phi_\mathrm{G}},
V_\mathrm{z_\mathrm{G}})=(-11.10,250.24,7.25)\,{\rm km}\cdot{\rm s}^{-1}$ \citep{Sc10,Sc12,IB21}.

The left-hand panel clearly shows the metallicity gradient as a function of the Galactic height,  $z_\mathrm{G}$. For high $|z_\mathrm{G}|\gtrsim 5\,{\rm kpc}$, the sample is dominated by metal-poor RR Lyrae ($\mathrm{[Fe/H]}$ between about $-1.7$ and $-1.5$) that belong to the Stellar Halo. Around $|z_\mathrm{G}|=3\,{\rm kpc}$, the mean metallicity starts to increase from $\mathrm{[Fe/H]}\approx -1.3$ to solar-like values, $\mathrm{[Fe/H]}\approx-0.5$, very close to the Galactic plane, at $|z_\mathrm{G}|<0.5\,{\rm kpc}$. Part of the population in this region is not visible due to the Galactic extinction and, therefore, it is larger than shown in the figure. The flattened metal-rich component extends all over the Galactic disc from the inner parts (at galactocentric radii $R_\mathrm{G}\approx 1\,{\rm kpc}$) to the very outer disc ($R_\mathrm{G}\approx 25\,{\rm kpc}$). It is important to note that the Stellar Halo RR Lyrae are also present at low Galactic heights (see, e.g., \citealt{IB21}); therefore, the gradient of the mean metallicity is in part driven by the evolution of the halo-to-disc fraction \citep{IB21}.

As may be seen from the middle panel in Figure \ref{fig:gaiadr3}, the most metal-rich RR Lyrae stars have a relatively thin vertical distribution consistent with an exponential model with a vertical scale height of $\approx 0.5\,{\rm kpc}$. This value is roughly consistent with the vertical scale height of the  Galactic stellar Thin Disc of $0.3$~--~$0.5\,{\rm kpc}$ \citep[e.g.,][]{Bovy2016}. The population with metallicities between $-1.2$ and $-0.8$ follows an exponential profile with a vertical scale height of $0.9$~--~$1\,{\rm kpc}$, consistent with the scale height of the Galactic Thick Disc \citep[e.g.][]{Juric2008}. At higher metallicities, the vertical profile rapidly broadens (scale height larger than $2\,{\rm kpc}$), setting the transition into the spheroidal Stellar Halo component.

The metal-rich RR Lyrae population not only occupies a space volume consistent with the Galactic Thin Disc but also shows cold kinematics (low velocity dispersion) typical of young stars in the Thin Disc. Since we expect that colder populations are also younger, the velocity dispersion can be used to estimate the population age \citep[see, e.g.,][]{Sharma20}. The right-hand panel shows the result of applying the \cite{Sharma20} model to a subsample of $4039$ GDR3 RR Lyrae with full available phase-space information. There is a clear age transition from the old ($\gtrsim 8$~--~$10\,{\rm Gyr}$) Halo/Thick Disc population (low $L_\mathrm{z}$ and $\mathrm{[Fe/H]}$) to the younger Thin Disc population ($\lesssim 7\,{\rm Gyr}$). However, we warn the reader that the absolute ages reported in Figure \ref{fig:gaiadr3} could be affected by the uncertainties that have not been considered in this simple analysis. Despite these uncertainties, one may see an age gradient reflecting the gradient in velocity dispersion as a function of the RR Lyrae metallicity.

In conclusion, this first examination of the RR Lyrae in GDR3 confirms the presence of a cold, Thin Disc-like population composed mostly of metal-rich RR Lyrae. The kinematics of this population is consistent with a typical young population in the Galactic Thin Disc, and it is challenging to explain such RR Lyrae through the classical formation scenario in which RR Lyrae must be old and metal-poor.

\subsection{Binary Metal-Rich RR Lyrae}
\label{sec:BinMetalRichIntro}

Binary stellar interactions can provide an alternative source of mass loss required to form young and metal-rich RR Lyrae. Such mass loss can be realised, for example, by material stripping by a binary companion through Roche-lobe overflow, e.g. \citet{Hurley02}, or tidal enhancement of wind mass loss, e.g. \citet{Tout88}. It should be remembered that a large fraction of non-interacting classical RR Lyrae are also expected to have a wide-orbital period binary companion, as we discuss in detail in Section~{\ref{sec:Method}}.

Systematic searches for RR Lyrae in binaries have produced more than 500 candidates using different methods: light travel-time effect \citep[e.g.][]{Li14,Hajdu15,Liska16a,Liska16,Ji18,Prudil19,Hajdu21}, astrometric anomalies \citep[e.g.][]{Kervella2019a,Kervella2019b}, radial velocities \citep[e.g.][]{Barnes21}, eclipses \citep[e.g.][]{Soszynski09,Richmond11,Soszynski11} or high-resolution images \citep[e.g.][]{Salinas20}. \citet{Li2023} recently focussed on the metal-rich RR Lyrae V838 Cyg, and found a long-period modulation (840 days) that can be related to binary interactions; see also \citet{Li2014}. However, the large majority of the candidates still lack a definite confirmation, and some systems have already been rejected \citep[see, e.g.][]{Salinas20}. Some of these methods, e.g. the astrometric anomaly searches \citep[e.g.][]{Kervella2019a,Kervella2019b}, are only sensitive to wide binaries on $100$~--~$1000\,{\rm AU}$ orbits. Such systems belong to the mentioned class of non-interacting (single-made) binary RR Lyrae discussed in detail in Section~\ref{sec:Method} and are of primary interest to this study.

Currently, there are only two robust confirmations of RR Lyrae in binary systems: the field star TU UMa \citep{Szeidl86,Saha90,Kiss95,Wade99,Liska16a} and the eclipsing variable OGLE-BLG-RR Lyrae YR-02792 in the Bulge \citep{Soszynski09,BEP,Smolec13}. Although the light curve of the Bulge star closely resembles that of classical RR Lyrae, it has been defined as an RR Lyrae \vir{impostor} since its low mass  ($0.26\,{\rm M}_\odot$) and some other pulsation properties are not consistent with the classic RR Lyrae stars \citep{Smolec13}. Given the peculiar low mass of this system, \cite{BEP} proposed that it was produced through mass transfer in the binary system and considered this star as the prototype of a new class of objects dubbed Binary Evolution Pulsators (BEP). In contrast, as we show in Section~\ref{sec:Result}, because of its large orbital period of $\approx 8000\,{\rm d}$, TU UMa belongs to the class of non-interacting (single-made) binary RR Lyrae.

Motivated by the BEP discovery, \citet{BEPoccurrence} made use of the population synthesis \texttt{STARTRACK} code \citep{STARTRACK} to model the population of binaries among RR Lyrae and Cepheid pulsators at solar-like metallicity ($Z_\odot=0.02$). They found that mass transfer allows stars to enter the IS through various routes that can be categorised into three main channels. In the dominant channel, the stripped donor star ignites helium and lands on the IS as an HB star\footnote{The fact that the HB channel dominates the RR Lyrae population in the \citet{BEPoccurrence} may not be evident directly from the study.~However, this may be readily appreciated from the bottom panel in Figure 5 of the paper, together with the fact that the panel shows a log-linear plot.}. In the less frequent cases, RGB or Asymptotic Giant Branch (AGB) stars cross the IS {\sl while} being stripped by the companion, leading to short-lived objects. The authors conclude that binary-produced RR Lyrae stars represent only a tiny fraction of the total RR Lyrae population ($<1 \%$). However, their estimated binary population of BEPs was not normalised by the total Galactic mass or the masses and metallicities of its components, as we discuss in more detail in Section~\ref{sec:Discussion}. In this paper, we model the dominant binary RR Lyrae formation channel proposed by \citet{BEPoccurrence} where the RR Lyrae stars form as previously stripped HB stars that land on the IS and argue that this channel can fully explain the Galactic population of metal-rich RR Lyrae.

\subsection{Present Study}

The dominant channel of binary RR Lyrae formation in the \citet{BEPoccurrence} study describes the stars stripped on the RGB by a binary companion so that, after stripping, they retain an appropriate amount of envelope to land in the IS. Binary RR Lyrae thus belong to a larger population of stars stripped during the RGB phase by their binary companions and retaining envelopes of different thicknesses. Within this larger population, the population of long-period hot composite subdwarf B (sdB) type stars is, presently, characterised most thoroughly. They are core-helium burning stars that have been fully stripped of hydrogen by a binary companion \citep{Heber16,Han2000, Han2002, Chen2013}.

Recently \citet{Vos2020} performed a population synthesis of hot composite sdB binaries with a detailed binary evolution code MESA \citep[Modules for Experiments in Stellar Astrophysics,][]{Paxton2011, Paxton2013, Paxton2015, Paxton2018, Paxton2019}. 
By modelling a population of binaries calibrated by the Galactic metallicity and star-formation history, \citet{Vos2020} have been able to {\sl simultaneously} reproduce the observed orbital periods, mass ratios, metallicities, and formation rates of observed hot composite sdB binaries \citep[][and references therein]{Vos2019} without explicitly fitting any parameters.

The \citet{Vos2020} simulations also include HB stars that have been partially stripped and could populate the IS. The population of partially stripped HB stars is, additionally, in agreement with the study by \citet{Xiong22}, where the authors identified a sub-population of blue large-amplitude pulsators (BLAPs) as products of the RGB mass transfer, and the study by \citet{LiTim22}, where the authors asteroseismologically observed a population of partially-stripped horizontal branch stars.

In this study, we use the same setup as in the \citet{Vos2020} simulations to predict the existence and investigate the properties of the Galactic population of binary RR Lyrae produced through binary mass stripping. We show, in particular, that RR Lyrae formed through binary interactions are common and likely constitute all the metal-rich Galactic RR Lyrae throughout the Galaxy and even dominate the RR Lyrae population in the Solar Neighbourhood. The analysis presented in this work is also intended for the construction of the observationally-constrained binary population synthesis sample.

We summarise the population setup and the RR Lyrae formation scenario in Section~\ref{sec:Method}. We further present the observational properties of the binary RR Lyrae population in Section~\ref{sec:Result}, arguing that they provide a natural explanation for the Galactic population of metal-rich RR Lyrae consistent with the current observations. Finally, we discuss the implications of our modelling in Section~\ref{sec:Discussion}.

\section{Binary and Single Evolution Modelling}
\label{sec:Method}

Subsequently, throughout the paper, we use the following definitions:
\begin{itemize}
    \item Truly single RR Lyrae: RR Lyrae that are truly single and formed through single stellar evolution.  We also call them single-made single RR Lyrae.
    \item Single-made binary RR Lyrae: Systems in which the RR Lyrae progenitor underwent single stellar evolution while having a binary companion. The companion, situated on a sufficiently distant orbit, did not influence the RR Lyrae formation process.
    \item Binary-made RR Lyrae: RR Lyrae that formed through interactions with a companion.
\end{itemize}
Single-made RR Lyrae (both truly single and binary) are said to form through the Single Evolution Channel, whereas binary-made RR Lyrae are said to form via the Binary Evolution Channel. While all three RR Lyrae groups are studied herein, a particular emphasis is placed on the latter two categories, which pertain to binary RR Lyrae\footnote{It is implicit that such binary RR Lyrae might also have distant triple or higher-order multiple companions \citep{Toonen2016}}.

\subsection{Formation Scenario}

\begin{figure*}
    \centering
    \includegraphics[width=\linewidth]{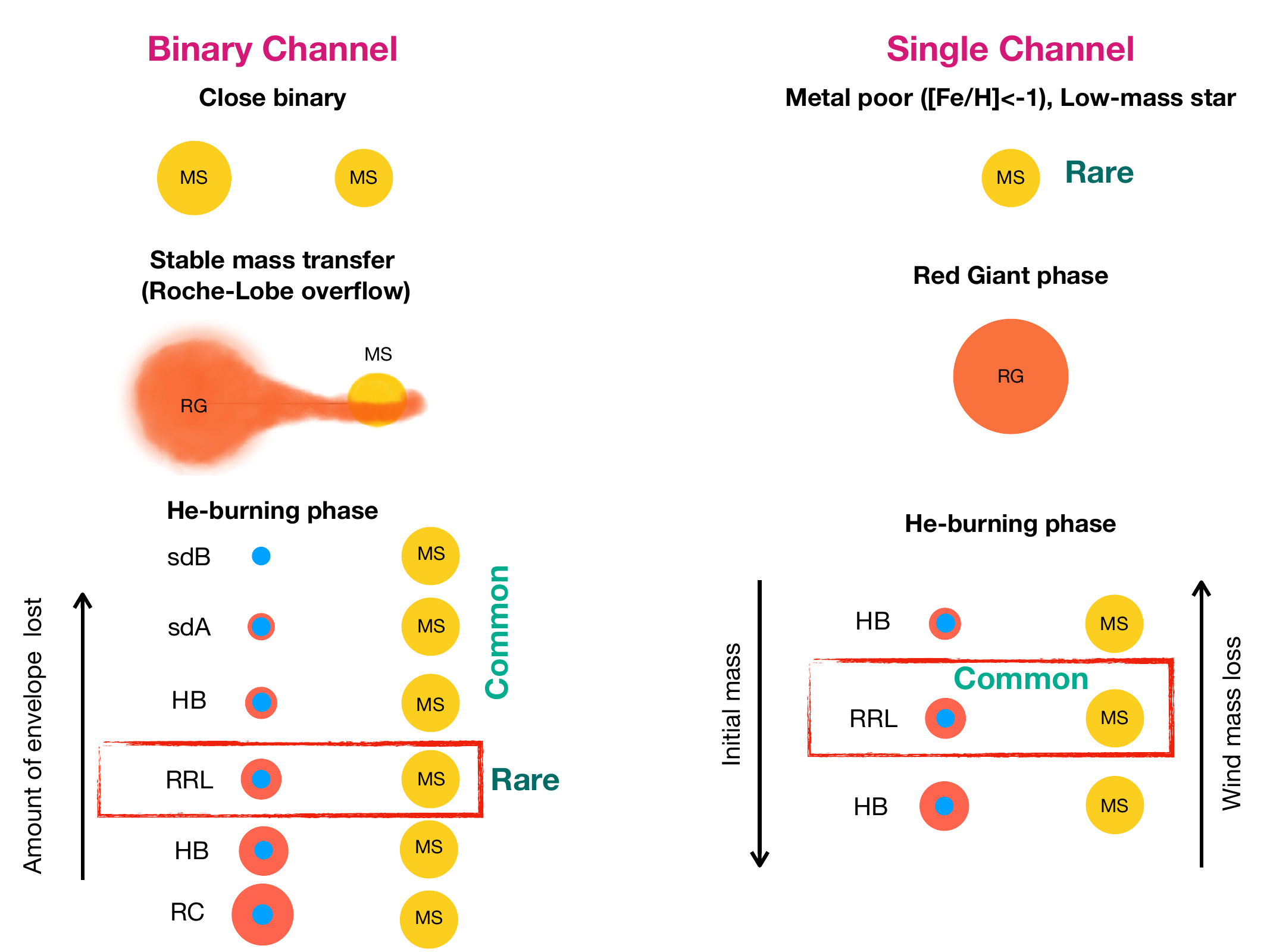}
    \caption{
    Sketch representing the two main formation channels for RR Lyrae stars. In the Binary Evolution Channel,  a typical main-sequence star (MS) gets stripped by a companion and ignites helium in its core (a common process). 
    If the red giant is stripped of only a fraction of its envelope, it produces a horizontal branch star (HB, common process). Depending on the amount of mass loss, a fraction of such HB stars will fall in the instability strip producing RR Lyrae stars (a rare outcome). In the Single Evolution Channel, the fraction of HB stars entering the instability strip depends on the metallicity, the initial mass of the star, and the wind mass loss during the Red Giant phase. Production of RR Lyrae stars through the Single Evolution Channel is common for old, metal-poor stars ($\mathrm{[Fe/H]}< -1$, a subdominant/rare population in the Milky Way), but very rare or almost impossible for metal-rich stars (see text for additional details).
    }
    \label{fig:sketch}
\end{figure*}

We summarise the two dominant formation channels for RR Lyrae in Figure~\ref{fig:sketch}. In the Binary Evolution Channel, which leads to binary-made RR Lyrae, the progenitor is a close binary ($100\, {\rm d} \lesssim P_{\rm orb}\lesssim 700\,{\rm d}$) that consists of two Sun-like main sequence stars in the field, such that the more massive primary ignites helium degenerately through a helium flash ($0.7\,{\rm M}_\odot \lesssim M_{\rm primary} \lesssim 2\,{\rm M}_\odot$). As the primary star evolves, it becomes a red giant and overflows its Roche lobe, non-conservatively transferring mass to the companion. If the red giant is sufficiently evolved by the time of Roche lobe overflow and has a sufficiently massive core, it ignites its core through a helium flash. Degenerate helium ignition drastically increases the core luminosity, thus leading to the contraction of the donor and terminating mass transfer. The stripped red giant remnant thus becomes an HB star similar to the HB stars produced from a Single Evolution Channel, albeit with a partly stripped envelope.

The exact dynamics of mass transfer is governed by the interplay of stellar evolution, the response of the donor to mass loss, the feedback of mass transfer on the orbital separation and the resulting effect on the degree by which the donor overflows its Roche lobe, and hence the mass transfer rate, e.g. \citet{Tauris2006,Bobrick2017,Vos2020}. In particular, the donor red giant tends to expand upon mass loss and contract after igniting helium. While we employ a detailed binary stellar evolution code to model the interplay of these processes, as we describe in Section~\ref{sec:mod}, we mention at this point that the resulting HB star can experience a wide range of the degrees of stripping and thus have a range of envelope masses.

The larger the fraction of the envelope that gets stripped, the more the He core of the HB star is exposed, and the bluer it becomes. In the limit of negligible stripping, the remnant is indistinguishable from HB stars from a Single Evolution Channel. In the limit of complete stripping, the naked helium-burning core remnant is observed as a hot sdB star. In the case of partial stripping, the surface temperatures may place the remnant in the instability strip. Since the HB star, apart from surface stripping, is otherwise similar to single HB stars, it will produce pulsations closely resembling those of classical RR Lyrae. Since the Binary Evolution Channel can operate at any metallicity, binary stripping allows metal-rich HB stars to acquire the thin envelopes they need to become RR Lyrae. 

In the Single Evolution Channel, which leads to single-made binary or single RR Lyrae, the progenitors are metal-poor ($\mathrm{[Fe/H]} \lesssim -1$) old stars ($t>10\,{\rm Gyr}$). They evolve faster than their metal-rich counterparts due to their low metallicity. Since the helium core mass in HB stars is always about $0.4$~--~$0.5\,{\rm M}_\odot$, such old, metal-poor stars have the lowest-mass envelopes. The combination of these effects places such single stars on the IS.

The Single Evolution Channel RR Lyrae stars represent a common product of single stellar evolution in metal-poor and old environments such as the Galactic Thick Disc and the Galactic Stellar Halo. However, stars in these MW components are rare and account for only a small portion of the total stellar content of our Galaxy (see Table~\ref{tab:Besancon}).
In the Binary Evolution Channel, a more abundant population of the Thin Disc field binaries experiences a  relatively common process of binary mass transfer, which may produce a binary-made RR Lyrae in the rare case when the stripping places the remnant in the IS.
 
Subsequently, we show that these two processes, with Rare$\times$Common and Common$\times$Rare probabilities, produce present-day RR Lyrae at comparable rates (within an order of magnitude). Since the Single and Binary Evolution Channels operate, respectively, at low and all metallicities, the combined action of both channels, therefore, can simultaneously explain the populations of metal-poor and metal-rich RR Lyrae.

\subsection{Single and Binary Modelling} \label{sec:mod}

\begin{table}
\begin{center}
\begin{tabular}{ |c|c|c|c|} 
 \hline
 Galactic bin & Age\,Gyr & Mass fraction & $[{\rm Fe}/{\rm H}]$ \\ 
 \hline
 Thin Disc - Bin 1 &$0-0.15$ & $0.030$ & $0.01\pm 0.12$\\
 Thin Disc - Bin 2 &$0.15-1$ & $0.069$ & $0.03\pm 0.12$\\ 
 Thin Disc - Bin 3 &$1-2$ & $0.076$ & $0.03\pm 0.10$\\ 
 Thin Disc - Bin 4 &$2-3$ & $0.072$ & $0.01\pm 0.11$\\ 
 Thin Disc - Bin 5 &$3-5$ & $0.132$ & $-0.07\pm 0.18$\\ 
 Thin Disc - Bin 6 &$5-7$ & $0.126$ & $-0.14\pm 0.17$\\ 
 Thin Disc - Bin 7 &$7-10$ & $0.171$ & $-0.37\pm 0.20$\\ 
 Bulge & $8-10$ & $0.192$ & $0.00\pm 0.40$\\
 Thick Disc & $10$ & $0.123$ & $-0.78\pm 0.30$\\ 
 Halo& $14$ & $0.008$ & $-1.78\pm 0.50$\\ 
 \hline
\end{tabular}
\caption{The Galactic bins, their ages, mass fractions, and metallicities, as used in this study. The stellar mass of the Galaxy was normalised to $6.43\cdot 10^{10}\,{\rm M}_\odot$ \citep{McMillan2011}. Star formation rates were considered uniform within each bin. The metallicities  $[{\rm Fe}/{\rm H}]$ were drawn from a Gaussian, with means and standard deviations provided herein.}
\label{tab:Besancon}
\end{center}
\end{table}

\begin{table*}
\begin{center}
\resizebox{\linewidth}{!}{%
\begin{tabular}{ |c|c|c|l|} 
 \hline
 Property & Functional Form & Parameter Range & Comments and references\\ 
 \hline
 IMF & ${\rm d} N/{\rm d} M_{\star}\propto M_{\star}^{-\alpha}$ & $\alpha=\left\{\def\arraystretch{1.2}\begin{tabular}{@{}l@{\quad}l@{}} $1.3$ & for $0.09\,{\rm M}_\odot < M_\star < 0.5\,{\rm M}_\odot$\\
 $1.8$ & for $0.5\,{\rm M}_\odot < M_\star < 1.53\,{\rm M}_\odot$\\
$3.2$ & for $1.53\,{\rm M}_\odot < M_\star < 150\,{\rm M}_\odot$\\
\end{tabular}\right.$  & \newline \begin{tabular}{@{}l@{}} Kroupa \& Haywood v6 model \\ Continuous, normalised \citep{Czekaj2014} \\ \citep{Kroupa2008,Haywood1997} \end{tabular} \\
 $M_{\rm primary, simulated}$ & --- & $0.7$~--~$2.1\,{\rm M}_\odot$ & All degenerately-igniting primaries\\
 $q_{\rm init}\equiv \dfrac{M_{\rm primary}}{M_{\rm secondary}}$ & ${\rm d} N_{\rm binary}/{\rm d} q_{\rm init}^{-1} \propto 1$ & $0<q_{\rm init}^{-1} < 1$ & \citep{Raghavan2010} \\
$q_{\rm init, binary-made}$ & --- & $1<q_{\rm init} < 3$ & All stably transferring binaries \\
 $P_{\rm orb}$ &  $\dfrac{\dd P_{\rm orb}}{\dd \log P_{\rm orb}}\propto 1$ & $1 < P_{\rm orb} < 10^4\,{\rm d}$ & Close binaries \citep{Abt1983}\\
$P_{\rm orb, binary-made}$ &  --- & $100\,{\rm d}< P_{\rm orb} < 700\,{\rm d}$ & All degenerately-igniting interacting primaries\\
$a_{\rm orb, single-made}$ &  --- & $1.2\,a_{\rm RLO, max, RGB}< a_{\rm orb} < 2\cdot 10^4\,{\rm AU}$ & All non-interacting primaries \citep{Abt1983}\\
Metallicity & ${\rm [Fe/H]}\propto \mathcal{N} ({\rm [Fe/H]}_i, \sigma_{\rm [Fe/H],i})$ & --- & Galactic metallicity distribution, Table~\ref{tab:Besancon}\\
Binary prob-ty & $0.45$ & --- & Galactic binary fraction \citep{Abt1983}\\
Close binary prob-ty & $0.25, 0.40$ & --- &\newline \begin{tabular}{@{}l@{}} Close binary fraction at ${\rm [Fe/H]} \approx -0.2$ and\\ halo metallicity, respectively \citep{Moe2019}\end{tabular} \\
Age cut & --- & $-300\,{\rm Myr} < t_{\rm RGB tip} - t_{\rm now} < 700\,{\rm Myr}$ & All present-day core-He burning stars \\
Mass loss parameters & $\dot{M}_\mathrm{accretor}=   (1-\alpha -\beta -\delta)  |\dot{M}_\mathrm{lost}|\dagger$   & $\left\{\def\arraystretch{1.2}\begin{tabular}{@{}l@{\quad}l@{}} $\beta = 1~{\rm if}~{\rm over-spinning}~{\rm or}~\tau_{\rm acc}<\tau_{\rm K-H}$\\ $\beta = 0~{\rm otherwise}$\\ $\alpha=\gamma=\delta=0$~{\rm always} \end{tabular}\right.$   &  
\newline \begin{tabular}{@{}l@{}} Effectively fully non-conservative\\ When $\dot{M}\gtrsim 10^{-5}$~--~$10^{-6}\,{\rm M}_\odot/{\rm yr}$\\ Mass loss with $J_z$ of accretor\\ \citep{Tauris2006} \end{tabular} \\
\hline
\end{tabular}}
\caption{The list of the model parameters used in this study. The modelled properties are listed in the first column. The next columns show the functional forms and model parameter ranges used to initialise the lists of properties. The last column lists the comments and references for each of the properties. While pre-initialising the full range of parameters for the Galactic populations, we simulated only the systems that may lead to stripped horizontal branch stars. Some of the rows show the range of the parameters selected for simulation. In addition to these properties, we used the Galactic model shown in Table~\ref{tab:Besancon} to assign the Galactic bins and ages to synthesise the population. $\dagger$Parameters $\alpha$, $\beta$, $\gamma$, and $\delta$ represent various forms of mass lost: $\alpha$ represents the fraction lost from the donor in the form of direct fast wind, $\beta$ denotes the fraction of mass ejected from the vicinity of the accretor, and $\gamma$ and $\delta$ symbolise the angular momentum content and mass lost from a circumbinary coplanar toroid, respectively.}
\label{tab:InitStars}
\end{center}
\end{table*}

We set up the Galactic population of all the present-day HB stars closely following the setup used in \citet{Vos2020} and summarise only the most important aspects here.

We initialise the Galactic population using the Besançon model \citep{Robin2003}, a standard Galaxy model calibrated by large-scale photometric surveys. Within the model, the Galactic population of stars is split into 10 bins: 7 bins for the Thin Disc and 1 bin for the Thick Disc, Halo and the Bulge, each with its mass and metallicity distribution, as shown in Table~\ref{tab:Besancon}, while assigning solar values to the hydrogen and helium abundances. We normalise the total present-day stellar mass of the Galaxy to $M_{\rm MW}=6.43\cdot 10^{10}\,{\rm M}_\odot$ following \citet{McMillan2011}. The mass fractions of the Thin Disc have been obtained as in \citet{Vos2020} by integrating the density profiles from \citet{Robin2003} and using the recent measurements for present-day stellar masses of the Thick Disc and Bulge from \citet{Robin2012,Robin2014} and the Halo from \citet{Deason2019}. This way, by randomly assigning each simulated object a Galactic bin, formation time and metallicity, following the Besançon model, we can keep track of the Galactic mass represented by each simulated object.

As the next step, we initialise a collection of binaries and single stars for the detailed stellar evolution modelling. We summarise the parameters and functions used for the stellar populations in Table~\ref{tab:InitStars}. The masses of single stars and the primary stars in binaries are drawn from the Kroupa \& Haywood v6 initial mass function (IMF) model from \citet{Czekaj2014}, as used in the Besançon model (see also \citealt{Kroupa2008, Haywood1997}). Since we are interested in present-day RR Lyrae and since all helium-burning stars exist for up to about $100\,{\rm Myr}$, we only select the primaries that reach the tip of the RGB not earlier than $300\,{\rm Myr}$ ago and not more than $700\,{\rm Myr}$ in the future. Furthermore, we are only interested in binaries that may become core-helium burning stars. Therefore, to model binary-made RR Lyrae, we consider binaries with initial orbital periods between $100\,{\rm d}$ and $700\,{\rm d}$, i.e. binaries in which the primary stars have developed a sufficiently massive core needed for helium ignition by the time of mass transfer. Similarly, to model single-made binary RR Lyrae, we consider the binaries where the primaries do not overflow their Roche Lobe on the RGB.

Finally, we focus only on the binaries with the primary mass between $0.7\,{\rm M}_\odot$ and $2.1\,{\rm M}_\odot$ that ignite their cores degenerately. Such binaries lead to RR Lyrae that have core masses similar to the single-made RR Lyrae from the classical Single Evolution Channel, and the models of preceding mass transfer in such binaries agree closely with the present-day long-period hot composite sdB populations. In contrast, the core masses of the HB stars that enter the IS after non-degenerate ignition typically differ from the canonical mass, and their orbital and companion properties may be affected by the choice of the mass transfer model parameters. Thus, within the scope of this study, we focus on degenerately-igniting systems.

To model binary-made RR Lyrae, we draw the companion masses from a uniform distribution in $q^{-1}$ \citep{Raghavan2010}, where $q\equiv M_{\rm primary}/M_{\rm secondary}$ in the range between $1$ and $3$, which fully encloses the binaries for which mass transfer proceeds stably. In comparison, all the interacting binaries with $q>3$ experience a common envelope phase. In this case, the binary companion and the giant’s core spiral into a short-period orbit ($P_{\rm post-CE}\lesssim 5\,{\rm d}$). For such post-common envelope systems, the core gets stripped so much that no HB star (and hence no RR Lyrae) can be produced, consistently with the lack of observed HB stars with short-period companions, e.g. Vos et al. 2024 (in prep). To model single-made binary RR Lyrae, we draw from the full range of $q$. We draw the binary orbital periods from the suitable log-uniform distribution from \cite{Abt1983} for close binaries with periods between $1$ and $10^4\,{\rm d}$. We then select the binaries in the mentioned $100$~--~$700\,{\rm d}$ period range for detailed simulation. For binary-made RR Lyrae progenitors, we assumed constant zero eccentricities for all the simulated binaries since red giants typically circularise by tides by the time they begin to overflow the Roche lobe \citep{Vos2015}. Finally, to associate each single or binary star with the stellar mass it represents in the Galaxy, we used the total binary fraction of $0.45$ and the close binary fraction (the fraction of binaries that have their orbital periods in the range of $1$~--~$10^4\,{\rm d}$) equal to $0.25$, which corresponds to the mean metallicity of the field $[{\rm Fe}/{\rm H}]\approx -0.2$ (\citealt{Moe2019}, see also \citealt{Mazzola20} and \citealt{Li22MB}).

We model the Galactic population of binary-made RR Lyrae by simulating $2060$ potential RR Lyrae progenitors in the detailed binary evolution code MESA, version r10390. We employ a standard model of mass loss by the binaries, in which binaries are conservative until the accreting star reaches super-critical rotation \citep{Popham1991, Paczynski1991, Deschamps2013}, which happens typically at mass transfer rates of $10^{-5}$~--~$10^{-6}\,{\rm M}_\odot/{\rm yr}$. Once the accretor reaches super-critical rotation, mass transfer is assumed to proceed in a fully non-conservative manner, with the carried away specific angular momentum equal to that of the accretor. Furthermore, following \citet{Vos2020}, we note that at about the same accretion rates, accretion proceeds on timescales shorter than the Kelvin-Helmholtz timescale of the accretor. As a result, the accreted material cannot cool down effectively and expands, also contributing to the non-conservativeness of mass transfer. For example, see \citet{Kippenhahn1977, Pols1994, Toonen2012}. Apart from being established theoretically, e.g. \citet{Soberman1997, Paxton2015}, the non-conservativeness for such progenitor systems is strongly supported by the observations of long-period hot composite sdB binaries \citep{Vos2020}. At the same time, the theoretically expected angular momentum loss model is yet to be entirely constrained by observations. We discuss the possible effects of other angular momentum prescriptions in Section~\ref{sec:Discussion}. Finally, we use the Neural Network-Assisted Population Synthesis code NNAPS\footnote{\href{https://github.com/vosjo/nnaps}{NNaPS} is available on GitHub} to extract the stellar parameters from the MESA runs.

To compare the population of binary-made RR Lyrae to the population of single-made classical RR Lyrae from the Single Evolution Channel in the same Galactic model, we simulate a Galactic sample of $10 000$ single progenitors of present-day HB stars by interpolating MIST tracks \citep{Mista} using the population synthesis code \textssc{SEVN} (\citealt{Iorio22}; \citealt{Mapelli20,Spera19,Spera17}). The MIST stellar tracks have been computed using the detailed binary stellar evolution code MESA \citep{Mistb}, and therefore, the comparison between the Single and Binary Evolution Channels is self-consistent to a certain extent regarding the stellar evolution model. The consistency may not be perfect, however. In particular, while our metallicity and helium-abundance prescriptions, as well as the wind prescriptions, agree, MIST is based on MESA version 7503, while our binary simulations are based on MESA version r10390. As a result, we expect that some parameters, such as convection prescriptions and the mixing length parameter, are different in the models. Also, note that MIST tracks assume relatively weak wind mass loss on the RGB. We discuss the implication of these differences, as well as alternative single evolution models with a more realistic wind range in Section \ref{sec:winds}.

To model both single and binary single-made RR Lyrae, we assume that solar-like Halo stars have the same binary fraction of $0.45$ as the field stars and have a close binary fraction of $0.4$, suitable for low metallicities $\mathrm{[Fe/H]}\lesssim -1$ \citep{Moe2019}. A binary RR Lyrae is considered to be single-made if its progenitor does not overflow its Roche lobe when its radius reaches $1.2\,R_{\rm RG, max}$. In such a case, it will transfer mass to the companion neither on the RGB nor during the temporary expansion after the helium flash, e.g., \citet{Fainer2022}. We sampled the binary orbital separations in a log-uniform manner, up to the maximum binary separation for field stars of $2\cdot 10^4\,{\rm AU}$ \citep{Abt1983}. At the same time, the range of mass ratios is not limited at all. Since RR Lyrae in the Halo can form much more effectively through the Single Evolution Channel, we do not consider binary-made RR Lyrae in the Halo.

For the purpose of this work, we define the instability strip boundaries following \citet{BEPoccurrence}:
\begin{equation}
\begin{split}
& \log \left( \frac{T_\mathrm{red}}{\mathrm{K}} \right)  \ = -0.05 \log \left( \frac{L}{\mathrm{L}_\odot} \right) + 3.94 \\
& \log \left( \frac{T_\mathrm{blue}}{\mathrm{K}} \right) = -0.05  \log \left( \frac{L}{\mathrm{L}_\odot} \right) + 4.00
\end{split}
\label{eq:is}
\end{equation}

In the subsequent analysis, whenever appropriate, we consider both the Galactic single- and binary-made populations of binary RR Lyrae, as well as the metal-poor Halo and Thick Disc populations of truly single RR Lyrae.

\section{Results}
\label{sec:Result}

\subsection{Single Stellar Evolution} \label{sec:sse}

\begin{figure}
    \centering
    \includegraphics[width=\columnwidth]{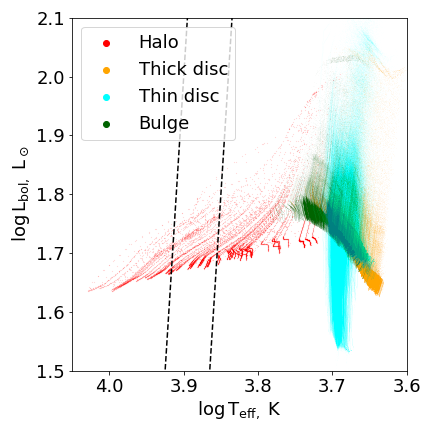}
    \caption{Evolution tracks for $10000$ single stars sampled from the Galactic model and evolved by interpolating MIST stellar tracks in the population synthesis code \textsc{SEVN} (see Section \ref{sec:mod}). Each point represents the properties of a given star at a certain time during the core-helium burning phase. Different colours indicate stars belonging to different Galactic components (Section \ref{sec:mod}): Stellar Halo (red, $57$ tracks), Thick Disc (orange, $865$ tracks), Thin Disc (cyan, $7904$ tracks), Bulge (dark green, $1174$ tracks). The black-dashed lines show red and blue limits of the instability strip (Equation~\ref{eq:is}). 
    }
    \label{fig:HRPhotSSE}
\end{figure}

In Figure~\ref{fig:HRPhotSSE}, we show the MIST evolution tracks of $10000$ stars during the core-helium-burning phase. The initial masses and metallicities of these stars are sampled from the Galactic model as described for single-made RR Lyrae progenitors (Section \ref{sec:mod}). Only old (${\rm Age}\approx 14\,{\rm Gyr}$), low-mass ($M_{\rm star} \approx 0.73\,\mathrm{M}_\odot$) and metal-poor ($\mathrm{[Fe/H]}\lesssim -1.5$) stars in the Stellar Halo (57 tracks) reach temperatures high enough ($T_\mathrm{eff} \gtrsim 7000\,{\rm K}$, $\log T_{\rm eff}/{\rm K}\gtrsim 3.85$) to enter the instability strip. Stars belonging to the other Galactic components -- Thick Disc (865 tracks), Bulge (1174 tracks), Thin Disc (7904 tracks) -- spend the core-helium burning phase in the red part of the HR diagram ($T_\mathrm{eff}\approx 5000\,{\rm K}$, $\log T_{\rm eff}/{\rm K}\approx 3.7$), forming the so-called red clump (RC). The Stellar Halo component is an efficient factory of RR Lyrae stars; about half ($53\,\%$) of stars belonging to the Halo component cross the IS during their core-helium-burning phase, while about a third ($28\,\%$) of stars spend almost all their core-helium-burning stage in the IS. 

The combination of the Galactic model with the MIST stellar tracks may seem to imply that no single-made RR Lyrae stars can be produced in any Galactic components other than Halo through single stellar evolution. This scenario is likely an oversimplification due to the uncertainties in the stellar evolution (e.g. stellar winds and chemical compositions, see Section \ref{sec:winds}) and Galactic properties. For example, in the Galactic population model we use \citep{Vos2020}, the stellar ages in the Bulge range between $8$ and $10\,{\rm Gyr}$ and the metallicity is higher than $-1.2$ at the $3\sigma$ level\footnote{The age of the stars in the Bulge are drawn from a uniform distribution between $8$ and $10\,{\rm Gyr}$, while the metallicity distribution is a Gaussian with centre and standard deviation set at $\mathrm{[Fe/H]}=0$ and $0.4\,{\rm dex}$, respectively.}. In contrast, \cite{Savino20} found that the Milky Way Bulge hosts metal-poor ($\mathrm{[Fe/H]}\lesssim-1.6$) and old RR Lyrae (${\rm Age}\gtrsim 13\,{\rm Gyr}$). The likely progenitors of these objects are not included in our Bulge sample. However, it is still unclear whether these RR Lyrae in the Bulge trace a clear separate Bulge component or just the inward extension of the stellar Halo (see, e.g. \citealt{Navarro21,Savino20,Villegas2017,Minniti99}). Moreover, even considering significant modifications in stellar evolution models (e.g. see Section \ref{sec:winds}, where we consider enhanced mass loss on the RGB), achieving RR Lyrae formation efficiency is challenging. This is consistent with the observations of RR Lyrae in other Galactic components, and especially considering the youngest populations (${\rm Age}\lesssim 8\,{\rm Gyr}$).

In conclusion, a clear bimodality emerges: in the Stellar Halo, the RR Lyrae stars are efficiently produced as a natural consequence of single stellar evolution. However, in the other Galactic components, the RR Lyrae production through single stellar evolution is inefficient or even absent. In such cases, it has to rely on enhanced mass loss or peculiar chemical compositions and generally (see Section \ref{sec:winds}) no plausible single stellar evolution models are capable of producing young RR Lyrae stars (${\rm Age}< 7\,{\rm Gyr}$). It is also worth mentioning that the in-situ part of the Stellar Halo \citep{Belokurov2020,Aurora} is more metal-rich than the rest of the Stellar Halo (see e.g. \citealt{IB21}). Hence, the alternative formation channels could have played a role in forming part of the RR Lyrae and also within the in-situ Halo.

\subsection{Physical Properties of Binary RR Lyrae} \label{sec:binprop}

\begin{figure}
    \centering
    \includegraphics[width=\columnwidth]{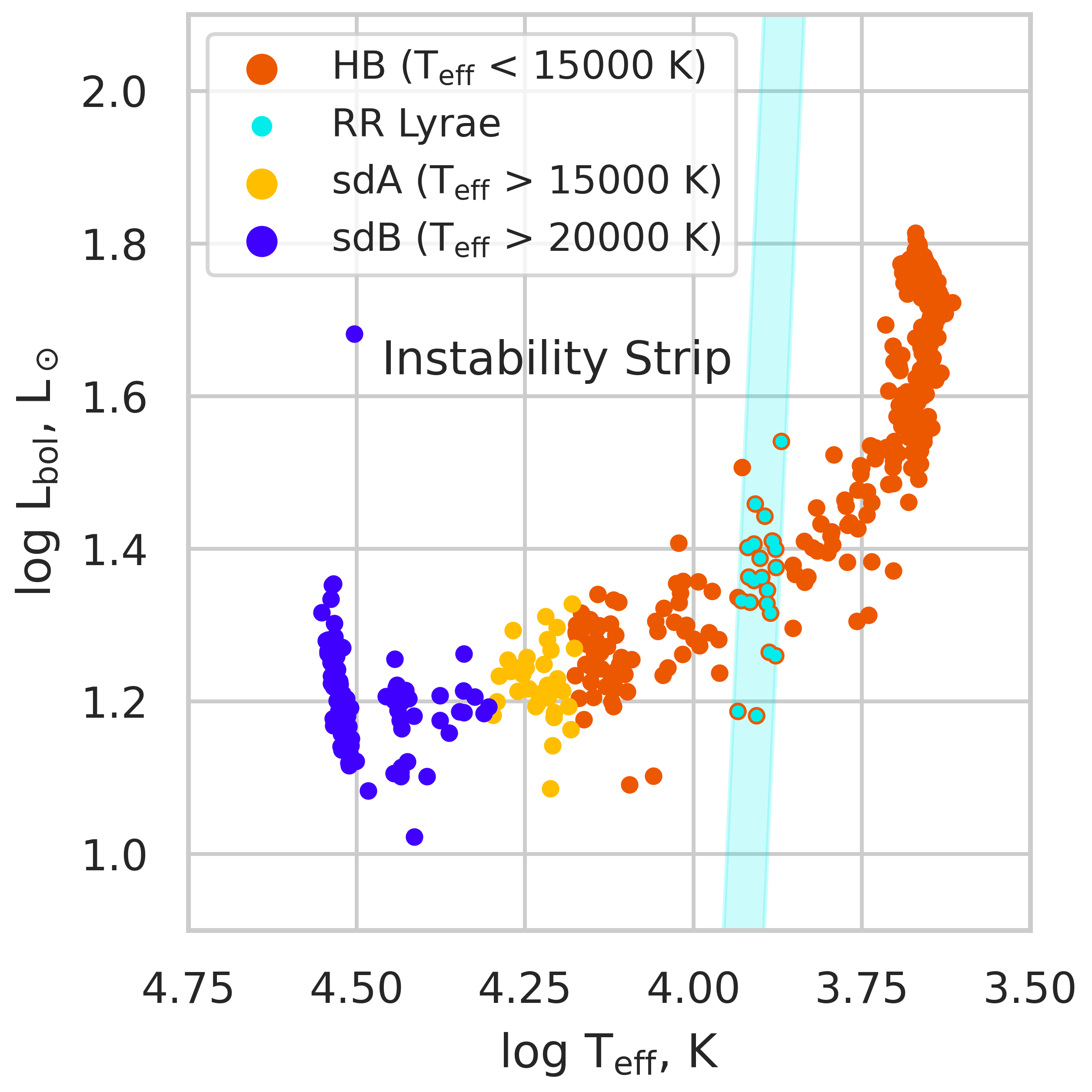}
    \caption{
    The HR diagram (bolometric luminosity versus effective temperature), based on our MESA models, shows the stars that have been stripped by their binary companions on the red giant branch (RGB) and undergo degenerate helium ignition: completely stripped subdwarf B stars (sdBs), nearly completely stripped 
    subdwarf A stars (sdAs), partially stripped HB stars, among which 22 objects are in the RR Lyrae instability strip. 
    }
    \label{fig:HRLTeff}
\end{figure}

\begin{figure}
    \centering
    \includegraphics[width=\columnwidth]{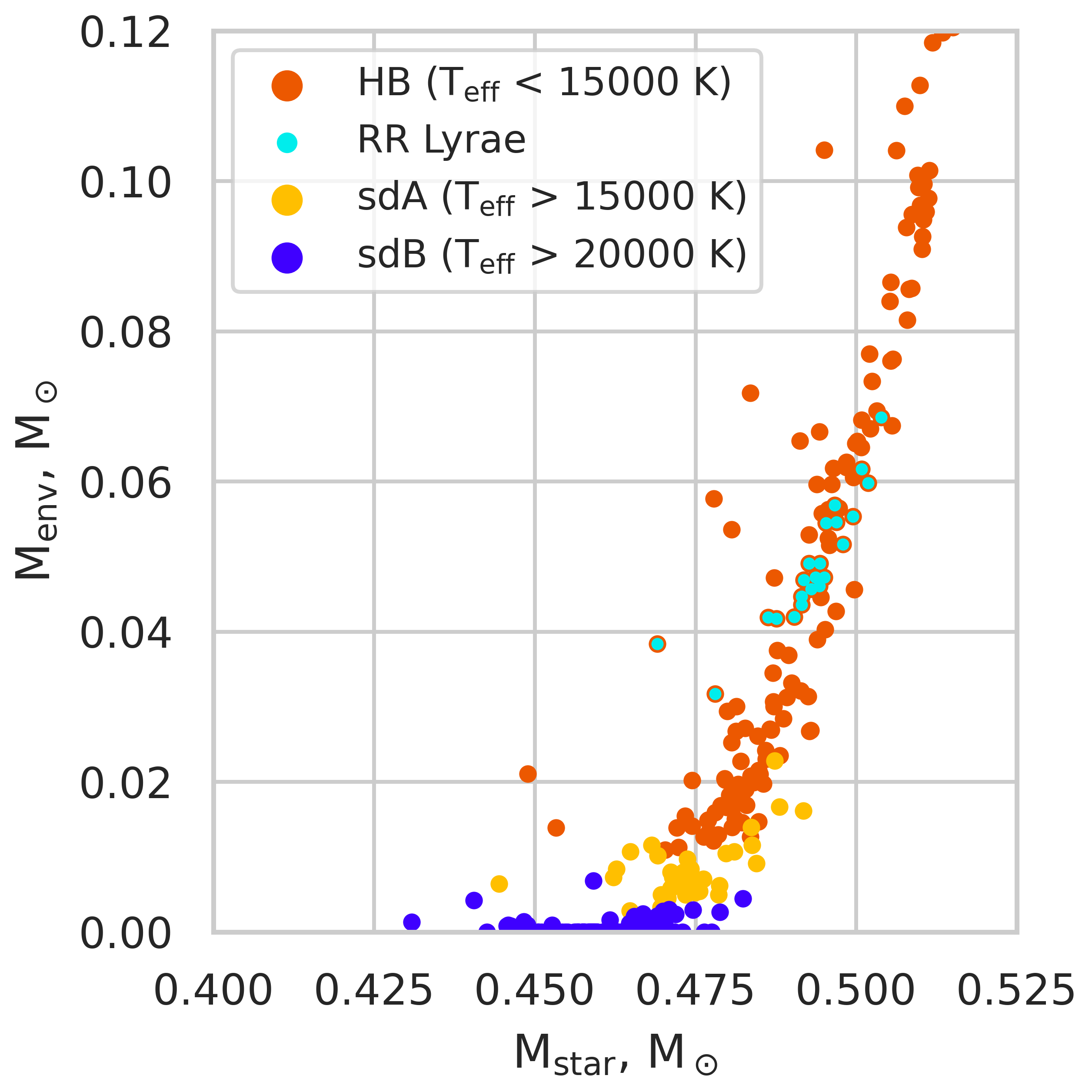}
    \caption{
    Same as Figure~\ref{fig:HRLTeff}, but showing the hydrogen envelope mass versus the total stellar mass. 
    }
    \label{fig:MStarMEnv}
\end{figure}

\begin{figure}
    \centering
    \includegraphics[width=\columnwidth]{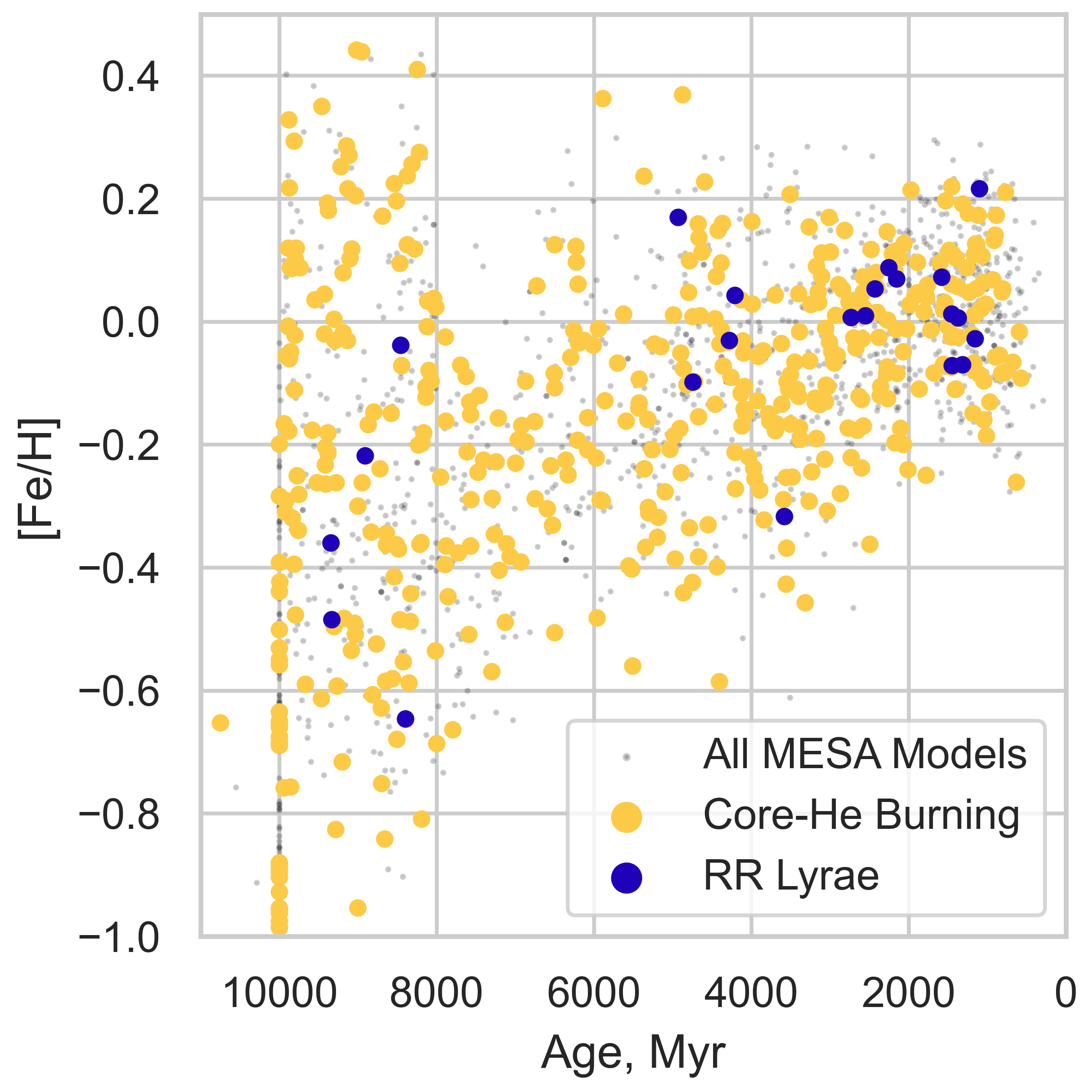}
    \caption{Age-metallicity relation for the binaries in our synthetic population. The RR Lyrae are marked with dark blue points, the core-helium burning stars are marked with yellow, and the systems that do not ignite helium are marked with grey.
    }
    \label{fig:AgeFeH}
\end{figure}

\begin{figure}
    \centering
    \includegraphics[width=\columnwidth]{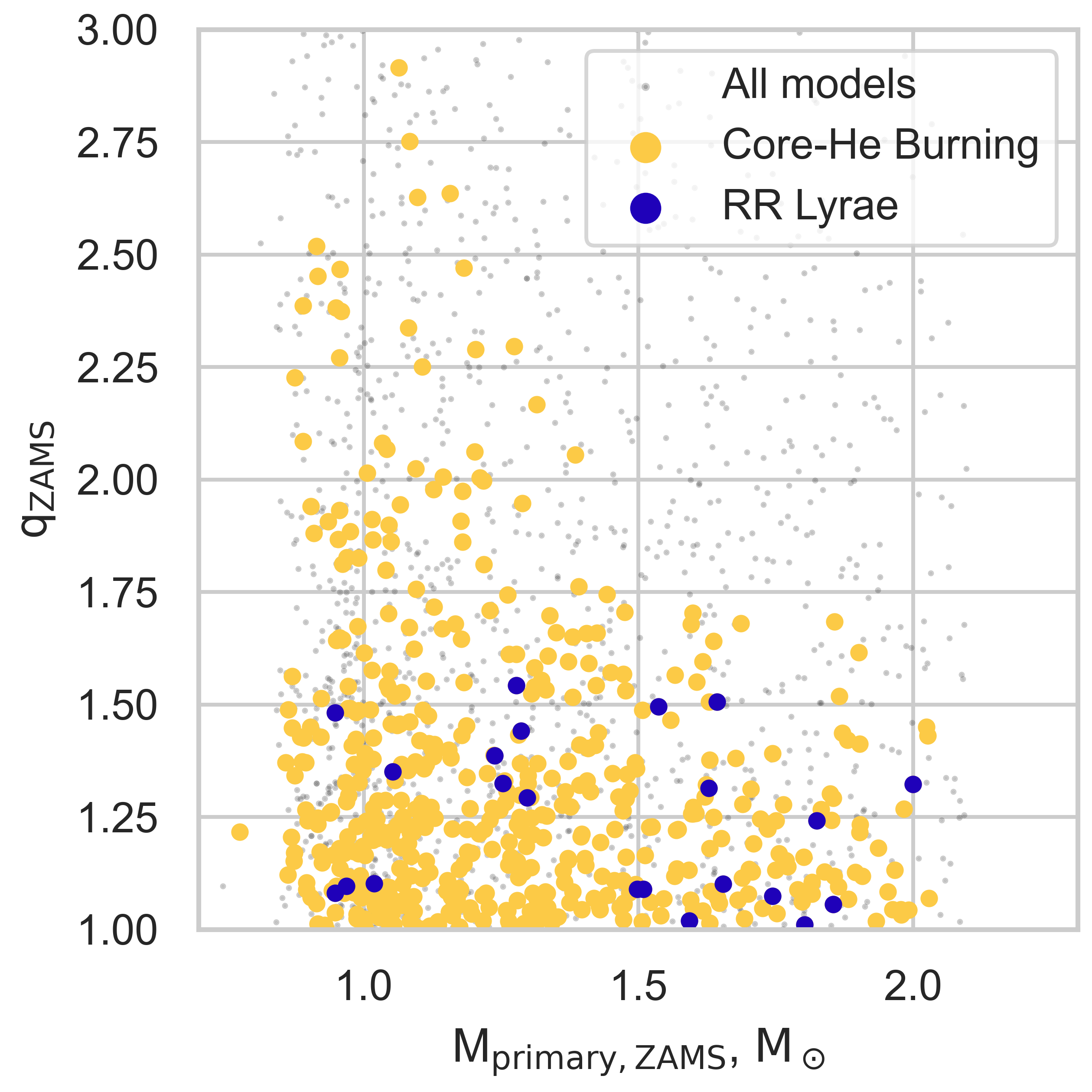}
    \caption{
    Properties of the progenitors for the binaries in our synthetic population. The horizontal axis shows the initial masses of the primaries (the stars that were initially more massive) in our binary dataset. The vertical axis shows the binary mass ratios $q_{\rm ZAMS}\equiv M_{\rm primary}/M_{\rm secondary}$. The RR Lyrae progenitors are marked with dark blue points, the core-helium burning star progenitors are marked with yellow, and the systems that did not ignite helium are marked with grey. 
    }
    \label{fig:ProgMQ}
\end{figure}

\begin{figure}
    \centering
    \includegraphics[width=\columnwidth]{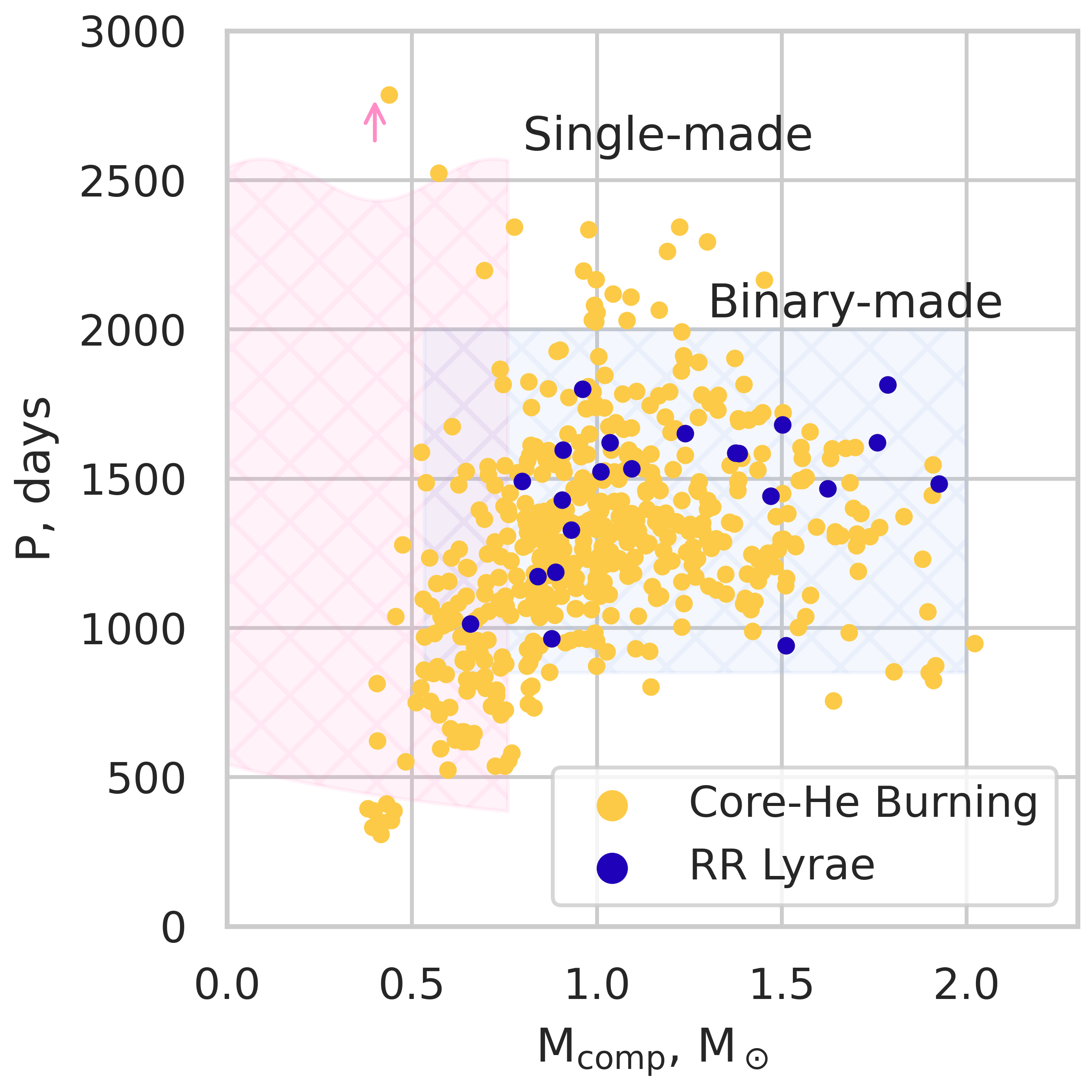}
    \caption{
    Orbital properties of the binaries in our synthetic populations containing a core-helium-burning star (yellow points). The RR Lyrae are marked with dark blue points. The horizontal axis shows the mass of the companion of the core-helium burning star. The vertical axis shows the binary orbital period. The shaded regions show the areas occupied by the binary RR Lyrae populations, which form either through binary mass stripping (Binary Evolution Channel, blue) or via evolution without binary interactions (Single Channel, pink). The pink arrow and the break in the pink-shaded area indicate that the distribution of the periods for the single channel RR Lyrae extends far outside of the plot range (up to $4\cdot10^9$ days).
    }
    \label{fig:MCompPeriod}
\end{figure}

In our models, 25\% (507 systems) of the simulated binaries produce a core-helium burning star that underwent a mass-stripping episode during the RGB evolution.

Depending on the initial parameters, such as orbital periods and mass ratios, binary companions strip different fractions of the donor envelopes. 
As shown in Figure~\ref{fig:HRLTeff}, the products of binary stripping occupy a well-defined place and form a continuum in the $L$~--~$T_{\rm eff}$ diagram. Complete stripping leads to hot ($T_{\rm eff}>20000\,{\rm K}$) naked helium-burning cores, observed as subdwarf B stars (sdBs, 122 objects). Near-complete stripping leads to colder subdwarf A stars (sdAs, 32 objects), whose envelope reprocesses the core radiation, making it cooler and redder ($15000\,{\rm K}<T_{\rm eff}<20000\,{\rm K}$). Partial stripping produces a continuum of red and cool stars observable as horizontal branch stars (HBs, 353 objects). A fraction of these HB stars lie on the instability strip and is observed as pulsating RR Lyrae stars (22 objects).

The binary-made RR Lyrae in Figure~\ref{fig:HRLTeff} in our sample occupy a range of luminosities between $15$ and $35\,{\rm L}_\odot$, which is similar to but systematically fainter than the classical (binary or single) single-made RR Lyrae from the Single Evolution Channel, which have luminosities closer to $45$~--~$50\,{\rm L}_\odot$ (e.g., see Figure~\ref{fig:HRPhotSSE}). Since the cores of RR Lyrae from the Single and Binary Evolution Channels are nearly identical, the fact that the metal-poor RR Lyrae are brighter than the metal-rich RR Lyrae from the Binary Evolution Channel can be seen as a natural consequence of their metallicity. Indeed, metal-poor stars have lower opacity, making it easier for the inner nuclear-burning regions to propagate the radiative energy to the stellar surface. Therefore, metal-poor stars are generally shorter-lived and more luminous than otherwise similar metal-rich stars \citep{Bazan1990,Kippenhahn2013}.

Structurally, the properties of binary-made RR Lyrae resemble those from the classical Single Evolution Channel. In Figure~\ref{fig:MStarMEnv}, we show the distribution of the envelope masses versus the total stellar mass for the sample of all the core-helium burning stars that ignited helium degenerately. The surface temperature of the stars correlates with the amount of hydrogen envelope left on the star, with the hottest sdB stars being nearly fully stripped of hydrogen ($M_{\rm sdB, env}\lesssim 0.007\,{\rm M}_\odot$) and the HB exhibiting a range of envelope masses, ranging from about $0.01\,{\rm M}_\odot$ up to approximately $1.0\,{\rm M}_\odot$. Because the majority of RR Lyrae have a narrow range of surface temperatures, they also have a relatively narrow range of envelope masses between $0.03\,{\rm M}_\odot$ and $0.07\,{\rm M}_\odot$. Furthermore, because the degenerate helium ignition occurs under roughly the same conditions, regardless of the progenitor mass, binary-made RR Lyrae have a narrow range of stellar masses between $0.47\,{\rm M}_\odot$ and $0.50\,{\rm M}_\odot$. Interestingly, our simulations show that degenerate core helium flash effectively triggers the termination of mass transfer in binaries. This occurs due to the rapid increase in luminosity caused by the helium flash, leading to the contraction of the donor star and effectively preventing any further mass transfer. Finally, as binary-made RR Lyrae have the same surface temperatures and core masses  as their classical single-made counterparts, we expect their pulsation properties to be similar, as we further discuss in Section~{\ref{sec:Discussion}}.

In contrast to the classical (single and binary) single-made RR Lyrae from the Single Evolution Channel, binary-made RR Lyrae have a broad range of ages and metallicities. As we show in Figure~\ref{fig:AgeFeH}, the ages and metallicities of binary-made RR Lyrae match those of intermediately young (ages between 1 and 10 Gyr) and metal-rich (${\rm [Fe/H]}>-0.8$) core helium-burning stars in the Galactic population. Therefore, most binary-made RR Lyrae trace the Galactic Thin Disc. Accordingly, metal-rich and young environments or dynamically cold orbits may be a good indicator for binary-made RR Lyrae. The apparent age gap between $5$ and $8\,{\rm Gyr}$ in the figure is likely related to the general decrease in RR Lyrae formation efficiency towards older ages and the effect of small-number statistics. Namely, only $3$ out of $20$ Thin Disc RR Lyrae have ages above $5\,{\rm Gyr}$. Similarly to RR Lyrae from the Single Evolution Channel, it follows from our simulations that binary-made RR Lyrae show an anti-correlation between their luminosity and metallicity, and the envelope mass and metallicity.

As we show in Figure~\ref{fig:ProgMQ}, binary-made RR Lyrae originate from binaries with primary masses in the range between $0.95$ and $2.0\,{\rm M}_\odot$ with mild mass ratios $q\equiv M_{\rm primary}/M_{\rm secondary}$ between $1.0$ and $1.55$. Binaries with higher mass ratios experience more extreme mass transfer, with enhanced stripping suppressing the production of RR Lyrae. Since no systems with $q>1.55$ produce RR Lyrae in our simulations, the fraction of such systems can be at most of order $1/22\approx 5\,\%$. The initial orbital periods of the progenitors are between $100$ and $700\,{\rm d}$, and these periods anti-correlate with the primary mass. The anti-correlation occurs because more massive stars evolve into smaller RGs and have shorter orbital periods at a given separation. We also have checked that all the systems we simulated fully enclose the parameter space of the binary-made RR Lyrae progenitors.

All binary-made RR Lyrae produced through the Binary Formation Channel have a binary companion. As we show in Figure~\ref{fig:MCompPeriod}, the present-day binary-made RR Lyrae have orbital periods between $1000$ and $1800\,{\rm d}$, and companion masses between $0.65$ and $1.9\,{\rm M}_\odot$. In other words, the binary orbital periods have increased compared to their initial values due to mass transfer. The range of companion masses may be easy to explain by recalling that binary-made RR Lyrae progenitors had primary masses between $0.95$ and $2.0\,{\rm M}_\odot$ and mild mass ratios between $1.0$ and $1.55$. Furthermore, there is a correlation between the initial primary mass and age, as the present-day RR Lyrae have recently ascended the RGB. Moreover, since the progenitor binaries initially had mild mass ratios, the present-day companion masses also correlate with age. In particular, only the systems younger than $3\,{\rm Gyr}$ have companion masses larger than $1\,{\rm M}_\odot$ and, conversely, all the systems older than $3\,{\rm Gyr}$ have companion masses lower than $1\,{\rm M}_\odot$. A similar correlation exists between the companion mass and the present-day metallicity, with all the systems bearing a $M_{\rm comp}\gtrsim 1\,{\rm M}_\odot$ companion having a solar-like metallicity above $\mathrm{[Fe/H]}\gtrsim -0.1$.

Many RR Lyrae from the Single Evolution Channel also have binary companions. In this case, however, the companion was on a sufficiently wide orbit, such that it did not perturb the single stellar evolution of the RR Lyrae progenitor. An example of such a system could be an old, metal-poor star (e.g. $0.8\,{\rm M}_\odot$, $\mathrm{[Fe/H]}=-1.5$) that had a far away companion at $10^3\,{\rm AU}$. In contrast to the Binary Evolution Channel, as we show in Figure~\ref{fig:MCompPeriod}, the orbital periods of such companions span a much wider range of $7$ orders of magnitude, from  $600\,{\rm d}$ to $4\cdot 10^{9}\,{\rm d}$. In particular, we estimate that only about $1$ in $12$ Halo-like metal-poor RR Lyrae in binaries will have a companion with an orbital period between $500$ and $2000\,{\rm d}$, and most of the companions will be on much wider orbits. In other words, the range of orbital periods that leads to stripping on the RGB is much narrower than the range of orbital periods needed for the secondary not to perturb the primary. 

Due to their low metallicities, the red giants from the Single Evolution Channel only expand to about $80$~--~$100\,{\rm R}_\odot$ \citep{Mista} and since binary interactions did not widen their orbits as in the Binary Evolution Channel, the orbital periods of such binaries can be as small as $600\,{\rm d}$. The companion mass range of single-made binary RR Lyrae from the Single Evolution Channel also differs from the Binary Evolution Channel. Since the Single Evolution Channel represents an old population, the MS companions that are more massive than the RR Lyrae progenitors have already turned into white dwarfs, while the MS companions less massive than the RR Lyrae progenitors have present-day masses below about $0.75\,{\rm M}_\odot$. Finally, a fraction of binary RR Lyrae from both Single and Binary Evolution Channels will have a wide-period tertiary or higher-order companion, e.g. \citet{Toonen2016}.

We present detailed properties of binary-made RR Lyrae in Table~\ref{tab:BMadeRRLDat} and those of single-made binary RR Lyrae in Table~\ref{tab:SMadeRRLDat}.

\subsection{Expected Observed Properties of Binary RR Lyrae}
\label{sec:propbin}

\begin{figure} 
    \centering
    \includegraphics[width=\columnwidth]{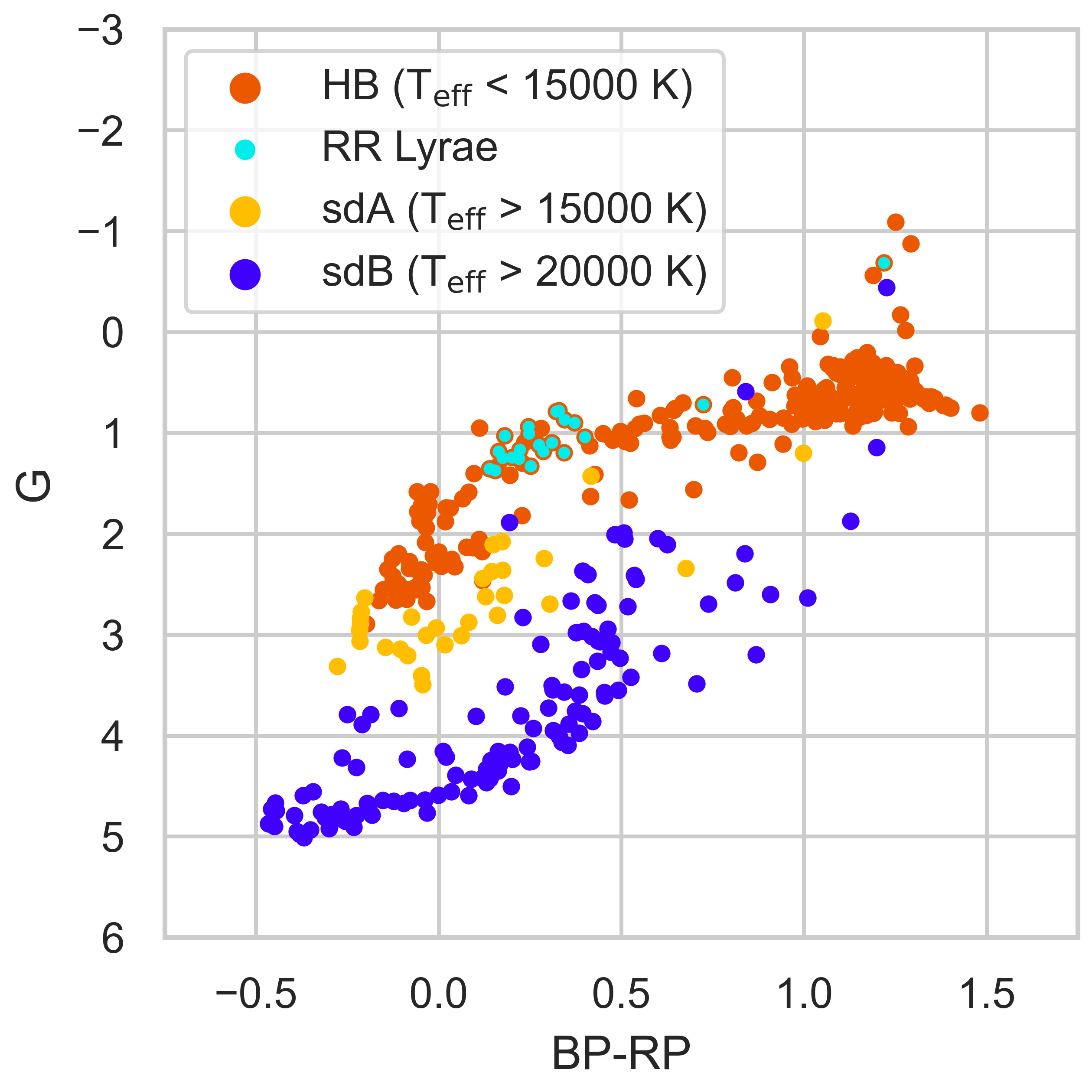}
    \caption{Same as Figure~\ref{fig:HRLTeff} but showing the HR diagram in the {\it Gaia} eDR3 bands (photometric G-band magnitudes and BP-RP colours).
    }
    \label{fig:HRPhot}
\end{figure}

\begin{figure}
    \centering
    \includegraphics[width=\columnwidth]{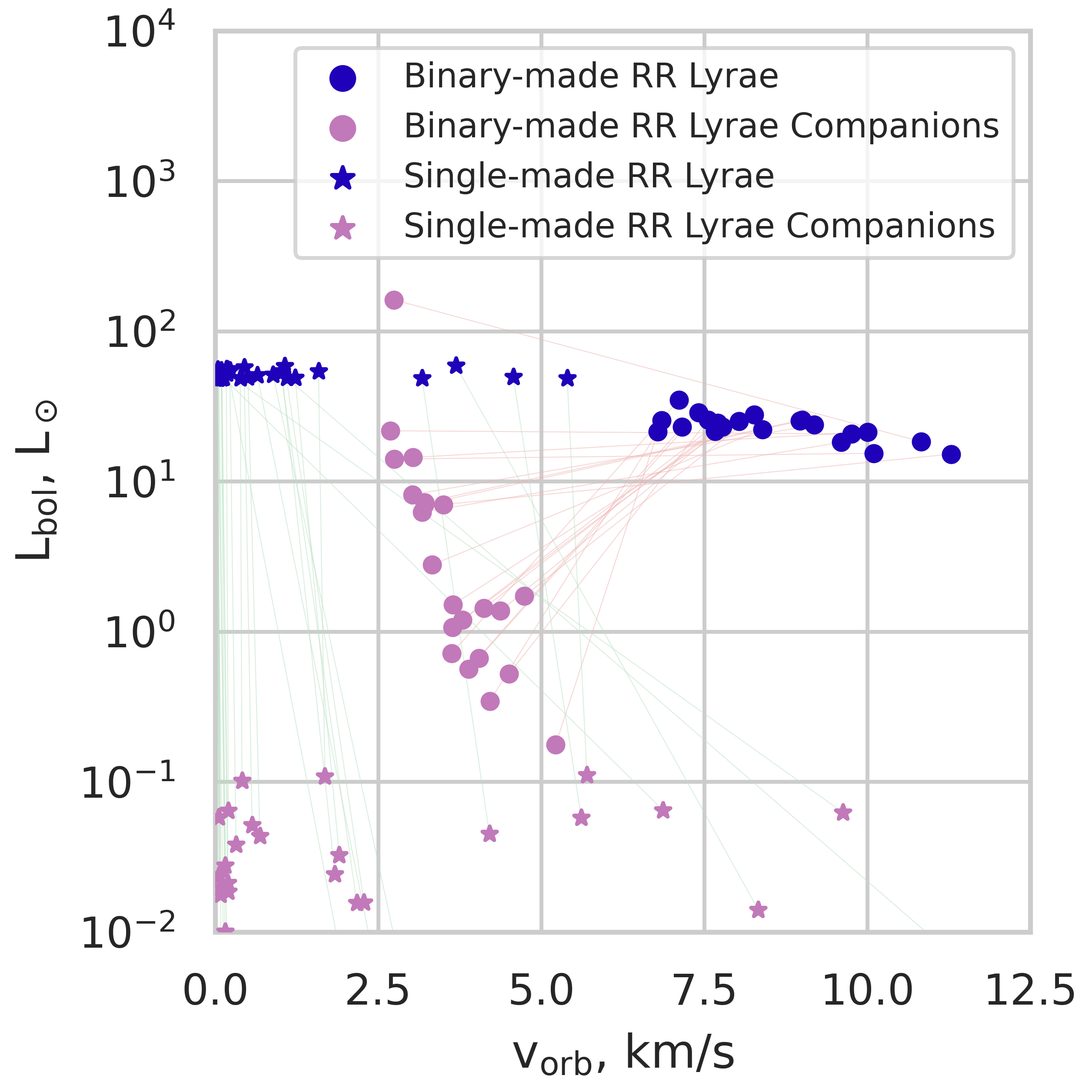}
    \caption{
    Properties of the binaries that host RR Lyrae stars in our synthetic population. The horizontal axis shows the orbital velocities of both components (blue markers for the RR Lyrae, pink markers for the companions), while the vertical axis shows their bolometric luminosities. The markers show the RR  Lyrae stars formed either through binary mass stripping (binary channel, circles) or through evolution without binary interactions (single channel, stars). The lines connect the stars belonging to a given binary system (red for the binary channel systems, green for the single channel systems).
    }
    \label{fig:VOrb}
\end{figure}

In Figure~\ref{fig:HRPhot}, we show a photometric HR diagram for stripped core-helium-burning stars, using the {\it Gaia} eDR3 G-band magnitude and BP-RP colours. Most binary-made RR Lyrae occupy a relatively narrow range of colours between $0.1$ and $0.4$ and G-band magnitudes between $1.4$ and $0.8$. The range of colours and magnitudes is consistent with the RR Lyrae observed by Gaia. In particular,  the range of magnitudes matches the observed metal-rich end of the luminosity-metallicity relation observed for the RR Lyrae in the field \citep{Muraveva2018,Garofalo2022,Li22}.

While metallicity is the primary differentiating property between the Single and Binary Evolution Channels of RR Lyrae, direct detection of the binary companions will provide important confirmation of either scenario. Moreover, as we discuss in Section~\ref{sec:Discussion}, measuring the properties of these companions may offer valuable new constraints for the binary evolution models. 

As may be seen from Figure~\ref{fig:HRPhot}, an 11  per cent fraction of binary-made RR Lyrae (2 out of 22 systems in the Figure) is located redwards from the main clump of RR Lyrae. These systems have an evolved companion. If the companion is not resolved in observations, it modifies the overall colour and G-band magnitude of the binaries, moving them closer to the RGB. Such systems are observationally interesting since, when unresolved, they exhibit RR Lyrae pulsations away from the IS. RR Lyrae from the Binary Evolution Channel may have evolved companions because even binaries with initial mass ratios close to unity may produce RR Lyrae. The orbital separations and the RR Lyrae properties in such systems are similar to the general population of binary-made RR Lyrae with main sequence companions.

Similarly, in the Single Evolution Channel, approximately $0.3$ per cent of all systems will have wide-orbit binary companions on the RGB. Such systems were born with near equal mass companions so that the secondary reached the RGB shortly after the primary, i.e. while the primary was still an RR Lyrae. Located away from the IS, RR Lyrae with evolved companions likely have not been included in the spectroscopic or photometric selection functions. On the other hand, detecting and characterising such systems will put tight constraints on their progenitors and thus will be extremely valuable. Finally, we predict that approximately $0.03$ per cent of metal-poor single-made RR Lyrae have a large-orbital-period binary companion which is also RR Lyrae. Such systems can only form in the Single Evolution Channel.

In Figure~\ref{fig:VOrb}, we show the orbital properties for the single- and binary-made binary RR Lyrae from the Single and Binary Evolution Channels. The properties of binary companions from both channels are notably different. The binary-made RR Lyrae in the metal-rich Binary Evolution Channel can have bright, close-by A, F, G and K-type companions. Indeed, such RR Lyrae have descended from relatively young binary progenitors and had companions with mild initial mass ratios. By now, all these binary-made RR Lyrae have been stripped to masses of $M_{\rm RRL}\approx 0.5\,{\rm M}_\odot$ while their companion masses have changed little. Therefore, such systems always have more massive, comparably bright present-day companions with orbital velocities of $v_{\rm orb, comp}\lesssim 5\,{\rm km}/{\rm s}$. Due to smaller masses, the orbital velocities of RR Lyrae in this channel are larger, of order $v_{\rm orb, RRL}\gtrsim 7.5\,{\rm km}/{\rm s}$. In the Single Evolution Channel, the companions, on the opposite, are always dim, $L_{\rm comp}\lesssim 0.1\,{\rm L}_\odot$, and less massive than RR Lyrae. However, since both stars have relatively low masses, $M_{\rm RRL/comp}\lesssim 0.75\,{\rm M}_\odot$, and due to a very broad { orbital} period distribution, the majority of the binaries will have vanishing orbital velocities $v_{\rm orb, RRL/comp}\lesssim 2\,{\rm km}/{\rm s}$. Due to the short-period tail of the orbital period distribution, about $20$ per cent of the binary systems from the Single Evolution Channel may reach orbital velocities of $v_{\rm orb, RRL/comp}\lesssim 10\,{\rm km}/{\rm s}$.

Constructing an observational sample of binary companions to binary- or single-made RR Lyrae requires one to overcome several challenges. Identifying composite RR Lyrae from a single spectrum is impossible for most systems because the companions are dimmer than RR Lyrae whilst not differing sufficiently in surface temperature. The systems with evolved companions on the RGB, shown in Figure~\ref{fig:HRPhot}, may be observed as composites. However, they need to be first identified as RR Lyrae despite their location away from the RR Lyrae clump on the colour-magnitude diagram. Likewise, because of their small orbital velocities (and yet smaller radial components of the orbital velocity) and because the RR Lyrae pulsations themselves lead to spectral velocity shifts of the order of $50\,{\rm km}/{\rm s}$ \citep{Gillet2019}, disentangling both spectral components of binary RR Lyrae from a mid- or high-resolution spectral series is also challenging. Additionally, such disentanglement requires spacing the spectral series over $2.7$~--~$4.8$ years to cover the complete orbits of the systems. 

If disentangled, similar to the observed companions of sdB binaries \citep{Vos2020}, non-convective companions of binary-made metal-rich RR Lyrae will be enriched with CNO products from the preceding accretion from an evolved red giant, the RR Lyrae progenitor. Due to their wide orbits, binary RR Lyrae are also good targets for {\it Gaia} DR3 and subsequent {\it Gaia} data releases providing orbital solutions, bearing in mind the possible caveat that the RR Lyrae variability may complicate the {\it Gaia} orbital solution. The binary orbital period distribution in the current {\it Gaia} DR3 dataset only reaches up to about $1000\,{\rm d}$ \citep{Gaiabinaryb}, which is the lower end of the expected orbital period distribution for both single- and binary-made binary RR Lyrae. Therefore, the current and future {\it Gaia} data releases may potentially probe the lower end of the orbital period distribution for binary RR Lyrae. With the above challenges in mind, the current most yielding source of binary RR Lyrae, perhaps, may still be the existing candidate search techniques mentioned in Section~\ref{sec:Introduction} that include the light time travel effect, astrometric anomalies, radial velocities, eclipses and imaging. Finally, one should remember that binary companions to RR Lyrae are expected both in the Thin Disc (the main location of binary-made systems) and in the Thick Disc and Halo (the main location of single-made binary systems).

\subsection{Galactic Orbits and Rates}

\label{sec:Rates}

\begin{table*}
\centering

\begin{tabular}{cccccc}
\hline
Type & Thin disc & Thick disc & Bulge & Halo & Total \\
\hline
${\rm R}_{\rm tot}$, ${\rm kyr}^{-1}$ & 0~:~0.51$\pm$0.11 & 0~:~$<$0.06 & 0~:~0.13$\pm$0.09~~~~ & ~~~~~~9.46$\pm$1.67~:~0 & 9.46$\pm$1.67~:~0.63$\pm$0.13 \\
${\rm F}_{\rm tot}$, ${\rm M}_\odot^{-1}$ & ~~~~~~~0~:~(1.1$\pm$0.3)$\cdot10^{-6}$ & 0~:~$<$1.3$\cdot 10^{-7}$ & ~~~0~:~(8.6$\pm$6.1)$\cdot10^{-7}$ & ~~~~(1.0$\pm$0.2)$\cdot10^{-3}$:~0 & (1.0$\pm$0.2)$\cdot10^{-3}$~:~(1.0$\pm$0.2)$\cdot 10^{-6}$ \\
${\rm N}_{\rm tot}$ & ~~~~0~:~48 000$\pm$11 000 & 0~:~$<$5300  &  0~:~10 500$\pm$7400 & 523 400$\pm$92500~:~0 & 523 400$\pm$92500~:~58 500$\pm$12500~~ \\
$n_{\rm loc}$, ${\rm kpc}^{-3}$ & 0~:~43.6$\pm$9.7 & 0~:~$<$0.9  & 0~:~0~~~~~~~~~ & ~~~~~~~~9.2$\pm$1.6~:~0 & ~~9.2$\pm$1.6~:~43.6$\pm$9.3\\
${\rm N}_{500\,{\rm pc}}$ & 0~:~13.2$\pm$3.0 & 0~:~$<$0.4  & 0~:~0~~~~~~~~~ & ~~~~~~~~4.8$\pm$0.8~:~0 & ~~4.8$\pm$0.8~:~13.2$\pm$2.8\\
${\rm N}_{1\,{\rm kpc}}$ & 0~:~70.5$\pm$15.8 & 0~:~$<$2.6  & 0~:~0~~~~~~~~~ & ~~~~~~38.4$\pm$6.8~:~0 & 38.4$\pm$6.8~:~70.5$\pm$15.0\\
\hline
\end{tabular}
\caption{The Galactic populations, formation efficiencies and rates of the RR Lyrae in the Single and Binary Evolution Channels, as follows from our models, shown with statistical Poisson uncertainties. The entries in the table are given in the following format: `Value for the Single Channel~:~Value for the Binary Channel'. The columns list the contributions from the main Galactic components: the Thin and Thick Discs, Bulge, Halo, and the Galactic Total. The rows show the present-day formation rate, the formation efficiency (number of systems over the total mass of the component), the total number of systems, the local number density, and the number of systems in the $500\,{\rm pc}$ and $1\,{\rm kpc}$ neighbourhood of the Sun, for each of the Galactic components. 
}
\label{tab:Rates}
\end{table*}

In Table~\ref{tab:Rates}, we present the Galactic population and formation rates of single- and binary-made RR Lyrae from the Single and Binary Evolution Channels shown in Figure~\ref{fig:sketch}, as follows from our modelling. At present, the Galaxy is predicted to contain $523~400$ and $58~500$ RR Lyrae from the Single and Binary Channels, respectively.

RR Lyrae from the Single Evolution Channel dominate the Galactic population, making about $90\,\%$ of the Galactic RR Lyrae. In the fiducial model, all RR Lyrae from the Single Evolution Channel are located in the Galactic Halo. However, as we discuss further in Section~\ref{sec:Discussion}, the RR Lyrae from the Single Evolution Channel may be distributed between the Halo and the Thick Disc depending on the prescription for wind mass loss on the RGB. Under a plausible variation of wind mass loss parameters, the Galactic number of single-made RR Lyrae varies by no more than about $50\,\%$. Finally, $38\,\%$ of the RR Lyrae from the Single Evolution Channel have a binary companion with a typically large orbital period. In such single-made binary RR Lyrae, the progenitor did not interact with the companion during the red giant phase, evolving effectively as a single star.

The binary-made RR Lyrae account for about $10\,\%$ of all the Galactic RR Lyrae. However, these systems comprise {\it all} the young and metal-rich RR Lyrae in our models. About $10 500$ out of $58 500$ binary-made RR Lyrae reside in the Bulge. This is because the turn-off stars in the Bulge have masses $M_{\rm TO, bulge}\gtrsim 1.0\,{\rm M}_\odot$ and require stripping to land in the IS. For similar reasons, in our models, the Binary Evolution Channel produces no RR Lyrae in the Thick Disc. 

In the Bulge area of the Galaxy, we expect to find both binary-made RR Lyrae from the Bulge component (and part of the Thin Disc close to the Galactic plane) and single-made RR Lyrae from the inward extension of the Halo. Considering the assumed Galactic model and the different RR Lyrae formation efficiencies (Table~\ref{tab:Rates}), we estimate that less than $5$~--~$6\,\%$ of the RR Lyrae in the innermost part of the Galaxy (within $3\,{\rm kpc}$ from the Galactic centre) belongs to the binary-made metal-rich population. This estimate is comparable with the fraction of the most metal-rich RR Lyrae found in the Bulge area ($\lesssim 10\,\%$, see e.g. \citealt{Savino20}). Similarly, in the Galactic Disc area, we expect to find a mix of binary-made and single RR Lyrae from the Thin Disc and the Halo.  We predict that the binary-made RR Lyrae from the Thin Disc account for $20$~--~$25\,\%$ of the RR Lyrae population within $3\,{\rm kpc}$ from the Galactic Disc. These numbers are consistent with the fraction of metal-rich thin-disc like RR Lyrae found by \cite{IB21} in the \gaia DR2 data ($\approx 27\,\%$ for Galactic heights lower than $3\,{\rm kpc}$). This fraction  increases up to $70\,\%$  in the solar neighbourhood ($500\,{\rm pc}$ radius around the Sun). The Binary Evolution Channel has such a significant local contribution because most of the systems from the Binary Evolution Channel reside in the Galactic Thin Disc, which is much more concentrated on the Galactic Disc than the Halo populated by the RR Lyrae from the Single Channel.

Considering the predicted age range for the binary-made metal-rich RR Lyrae (Figure~\ref{fig:AgeFeH}), we could expect to find some of them in the intermediate-age Open Clusters of the Milky Way (2--9 Gyr), yet none of them have been observed so far. 
However, given the low formation efficiency of binary-made RR Lyrae ($\sim$$10^{-6}\,{\rm M}^{-1}_\odot$, see Table~\ref{tab:Rates}) and assuming $\sim$$10^5\,{\rm M}_\odot$ as the total mass stored in all the observed intermediate-age Open Clusters ($\sim$$10^2$--$10^3\,{\rm M}_\odot$  average mass per cluster, \citealt{Piskunov2008},  $\sim$$10^3$ observed clusters, \citealt{CastroGinard22}, and 10\% fraction of clusters older than 1 Gyr, \citealt{Bossini2019}), our model is consistent with not observing any RR Lyrae in Open Clusters.
Similarly, we expect few or no RR Lyrae variables in the most metal-rich Globular Clusters (${\rm [Fe/H]}>-1$, $\sim10^5$--$10^6\,{\rm M}_\odot$ per cluster). Indeed, most observed metal-rich Globular Clusters are devoid of RR Lyrae (e.g. NGC 5927 \citealt{Liller83}; NGC 6352, \citealt{Alcaino71}; NGC 6838, \citealt{McCormac14}; NGC 6496, \citealt{Abbas15})
or contain few candidates (47 Tuc, \citealt{Bono08}; NGC 6304, \citealt{DeLee06}; NGC 6366, \citealt{Ferro08}; NGC 6367, \citealt{Escobar06}; NGC 6624, \citealt{Liller78}). The only metal-rich Globular Clusters containing several confirmed RR Lyrae members are NGC 6388 and  NGC 6441. However, the  RR Lyrae hosted in such clusters are peculiar objects with long pulsation periods, consistent with the metal-poor population of field RR Lyrae, and likely produced by He-enriched progenitors (\citealt{Pritzl00,Moehler2006,Yoon2008,AlonsoGarcia21}, but see also \citealt{Bhardwaj22}).

Altogether, the single-made RR Lyrae are consistent with the observed population of Classical RR Lyrae. At the same time, the binary-made RR Lyrae are compatible with the observations of RR Lyrae in the Thin Disc and the Bulge, and with the detections (or non-detections) of RR Lyrae in  Open Clusters and most metal-rich Globular Clusters.

\section{Discussion}
\label{sec:Discussion}

\begin{table*}
\begin{tabular}{lcccccc}
\hline
\multicolumn{1}{c}{Catalogue} & $N_\mathrm{match}$ & $N_\mathrm{clean}$ & $f_\mathrm{disc/halo}$ & $f_\mathrm{rich/poor}$ & $f_\mathrm{disc/halo,control}$ & $f_\mathrm{rich/poor,control}$ \\ \hline
RR Lyrae yrBinCan \citep{Liska16}                   & 68                  & 22                  & 0.24 (4:17)            & 0.50 (10:20)            & { 0.19 (10:53)}                           & { 0.20 (40:200)}                           \\               
\cite{Hajdu21}$\dagger$                       & { 52}                  &  { 0}                 &  { -}               & { 0 (0:3) }             & { 0.34 (14:41)}                           &    { 0.52 (59:114)}                         \\
\cite{Kervella2019a}                 & 139                 & 73                  & { 0.51 (23:45)}            & 0.27 (18:67)           & { 0.34 (25:73)}                            & { 0.16 (22:133)}                           \\
\cite{Kervella2019b}                    & 7                   & 3                   & 2 (2:1)                & 2 (2:1)                & { 0.8 (8:10)}                             & { 0.17 (16:95)}                          \\ 
\cite{Prudil19}$\dagger$                    & 8                   & 1                   & 0 (0:1)                & 0 (0:1)                & { 0.63 (5:8)}                         & { 0.43 (17:40)}                            \\ 
\hline
\end{tabular}
\caption{
Fraction of disc-to-halo and metal-rich-to-metal poor stars in catalogues obtained by cross-matching the {\it Gaia} DR2 RR Lyrae catalogue \citep{IB21} with catalogues of candidate RR Lyrae in binary systems: the online database RR Lyrae yrBinCan (Candidates for Binaries with an RR Lyrae Component database \href{https://rrlyrbincan.physics.muni.cz}{RR Lyrae yrBinCan}, \citealt{Liska16}); 
the \citet{Hajdu21} and \citet{Prudil19} catalogues of candidates based on the light-travel time effect; the \citet{Kervella2019a} and \citet{Kervella2019b} dataset based on the proper motion anomaly and proper motion pairs.
Columns two and three indicate the number of matched stars and the number of stars matched that are also in the clean catalogue used in \citet{IB21}. 
The fourth and fifth columns represent the disc-to-halo and metal-rich-to-metal-poor stars ratio considering the matched star, while the last two columns indicate the same ratios on a {\it Gaia} DR2  RR Lyrae control sample in the same Galactic regions sampled by the cross-matched catalogue (see text for details). The values in parentheses are the number of stars associated with different subsamples (disc/halo, rich/poor). $\dagger$ \citet{Hajdu21} reanalysed the stars in the \citet{Prudil19} sample.} 
\label{tab:ObsRates}
\end{table*}

\subsection{The Current Population of Binary RR Lyrae}
\label{sec:RRLBinaryCats}

In this study, we show that binary-made RR Lyrae presented in Figure~\ref{fig:sketch} provide a natural explanation for the young and metal-rich population of RR Lyrae in the Thin Disc and Bulge. These RR Lyrae are abundant in the Galaxy and form an inherent part of the population of binary stripped stars. Their surface temperatures lie inside the broad continuum of the temperatures occupied by the observed populations of stripped core-helium burning stars. Therefore, the presence of binary-made RR Lyrae is insensitive to stellar modelling. Furthermore, the formation rates of RR Lyrae from the Binary Evolution Channel are comparable to those from the classical Single Evolution Channel in the Galaxy overall and even dominate the rates in the solar neighbourhood. Detecting and characterising such binary-made RR Lyrae systems, as we describe in Section~\ref{sec:DiscussionImplications}, will lead to significant novel constraints on binary stellar evolution and, potentially, binary populations in the Galaxy.

The current strongest (although indirect) observational evidence for our modelled population of binary-made RR Lyrae lies in the large observed population of RR Lyrae in the metal-rich and Thin Disc environments (see e.g.~\citealt{IB21}), as mentioned in Section~\ref{sec:Introduction}, and the good agreement between our models and their observed photometric properties, metallicities and formation rates. It is important to note that, in the crowded regions of the Thin Disc, the presence of RR Lyrae impostors (e.g. artefacts or eclipsing binaries) could bias the population properties. \cite{IB21} found  an overall low level of contamination ($\lesssim 5 \%$) in the {\it Gaia} DR2 RR Lyrae catalogue. \cite{RRLG3} and \cite{DR3var} confirmed the same level of contamination in the {\it Gaia} DR3 RR Lyrae catalogue. Using a gold catalogue of robustly classified RRab stars, \cite{IB21} found that such contaminants are not introducing biases in the RR Lyrae population either in their intrinsic properties (e.g. period, metallicity) or in their phase-space distribution.

In Table~\ref{tab:ObsRates}, we show the result of cross-matching the {\it Gaia} DR2 RR Lyrae catalogue in \cite{IB21} with several catalogues of binary RR Lyrae candidates using a $1\,{\rm arcsec}$ matching window. We categorise the matched stars into subpopulations based on their kinematic classification (disc-like and halo-like, third column) and on their metallicity (metal-poor, ${\rm [Fe/H]}<-1$, metal-rich, ${\rm [Fe/H]}\ge-1$, fourth column) from \cite{IB21}. For each catalogue, we also produce a control sample, collecting for each matched binary RR Lyrae candidate all the RR Lyrae from {\it Gaia} DR2 within a window of $1\,{\rm degree}$ and with the parallaxes consistent within $50\,\%$. We then separate the control sample in the disc-halo, metal-rich-metal-poor subpopulations (last two columns). Despite numerous matches, only a small fraction of them belongs to the clean catalogue by \cite{IB21} for which we have a good metallicity estimate and a clear association to the Galactic disc. The \citet{Hajdu21} and \citet{Prudil19} samples cover an area of the bulge not extensively analysed by \cite{IB21}. Hence, almost all the matches with the clean catalogue come from the RR Lyrae yrBinCan database \citep{Liska16} and the \cite{Kervella2019a} catalogue based on proper motion anomalies. Within these two catalogues, there are hints that binary candidates could be more common in the disc and/or the metal-rich RR Lyrae population. A more detailed analysis of the \citet{Hajdu21} candidates, as summarised in Appendix~\ref{sec:BinRRLCandidates}, shows an adequate qualitative agreement with this picture. However, given the low statistics, it is not currently possible to derive a definite conclusion. 

We also tried to cross-match the \cite{IB21} dataset with the binary catalogues from {\it Gaia} DR3 \citep{Gaiabinarya,Gaiabinaryb}, but only four stars fall within the matching window ($1$ arcsec, considering the stars' proper motions). Considering the RR Lyrae catalogue from {\it Gaia} DR3 \citep{RRLG3}, we found 11 matches with the DR3 binary catalogues: 10 in the \texttt{nss\_two\_body\_orbit} table (consistent with orbital two-body solutions) and 1 in the \texttt{nss\_acceleration\_astro} (astrometric solutions consistent with non-linear proper motions). Among the 10 objects consistent with two-body orbit models: 4 are eclipsing binaries with periods twice the pulsation periods, and hence they are likely misclassified RRc RR Lyrae; 5 have spectroscopic solutions (SB1, see \citealt{Gaiabinaryb}) of which 4 have periods consistent with RR Lyrae pulsation, and thus they are likely spurious solutions biased by the intrinsic variability, the other one has a period of 14 days. The last candidate ({\it Gaia} DR3 ID 5239771349915805952) is an RRc with an orbital period of $2287\pm312\,{\rm days}$. It is in the region of the open cluster IC2602, but it has been classified as an interloper by \cite{Gutierrez2020}. In the same work, the authors estimate a solar-like metallicity  (${\rm [Fe/H]} = 0.04\pm0.02$) and $T_\mathrm{eff}=4599\pm32\,{\rm K}$. Even considering the higher $T_\mathrm{eff}$ estimate from {\it Gaia} DR3 ($\approx 5188\,{\rm K}$), the star is too cold to be in the instability strip (see e.g. Figure \ref{fig:HRPhotSSE}). Indeed, \cite{Samus17} classify the star as a BY Dra Variable (i.e. a variable main sequence star, see also \citealt{Yamashita2022}).

One may distinguish some of the binary RR Lyrae from the Binary and Single Evolution Channels based on their eccentricity distributions. Indeed, single-made binary RR Lyrae from the Single Evolution Channel preserve their birth eccentricity distribution, which is close to thermal (${\rm d} N/{\rm d} e=2e$). In comparison, the progenitors of binary-made RR Lyrae are likely circularised by mass transfer, potentially becoming circular and distinct from single-made binary RR Lyrae. It is worth noting that the closely related post-mass-transfer long-period sdB stars do commonly show mild eccentricities $e\lesssim 0.5$, which were suggested to be caused by the interaction with a circumbinary disc forming during mass transfer \citep{Vos2015}. Moreover, several other classes of post-mass-transfer binaries, including depleted binary post-AGB stars \citep{Oomen2018, Oomen2020}, blue stragglers, CEMP-s and Ba stars, e.g., \citet{Leiner2019}, also show non-zero eccentricities. Therefore, while we do not model the eccentricity evolution of binary-made RR Lyrae, we expect that RR Lyrae from both channels have non-zero eccentricity distributions. The distinguishing feature of the single-made binary RR Lyrae may be extreme eccentricities $e\gtrsim 0.6$ that may be impossible to reach within the Binary Evolution Channel, e.g. \citet{Vos2017}. In addition to these effects, even a low-mass triple companion may further modify the eccentricity of the inner binary.

Observationally confirming binary companions for such stars is challenging. As we show in Section~\ref{sec:Result}, for the majority of systems, the wide-orbit companions are relatively faint and have a similar temperature to RR Lyrae. Therefore, a several-year-long sequence of spectroscopic observations, similar in duration to the orbital periods of the systems, might be needed to disentangle the spectra and the RR Lyrae pulsations. One difficulty is that the pulsations themselves may significantly influence the spectra. In addition, the light travel time effect in RR Lyrae, as used, for instance, in \citet{Prudil19, Hajdu21}, may be used to identify new binary candidates, alongside the other methods mentioned in Section~\ref{sec:Introduction}. The most substantial contribution to the number of confirmed binary RR Lyrae might come from future {\it Gaia} observations, for which several year-long orbital periods are optimal for detection.

As we also discussed in Section~\ref{sec:Result}, we expect a nearly $10\,\%$ fraction of binary-made RR Lyrae to have an evolved companion so that the stars that are not photometrically in the IS can exhibit RR Lyrae pulsations (see Figure~\ref{fig:HRPhot}). Similar systems can form in $0.3\,\%$ of cases through the Single Evolution Channel, and, in $0.03\,\%$ of cases, the Single Evolution Channel may even form binaries where both components are RR Lyrae. When considering portions of the sky close to the disc ($|b|<10^\circ$) with a low level of extinction ($ebv<0.2$), we note that the fraction of RR Lyrae from the \cite{IB21} sample with extreme red colours (${\rm BP-RP}>1$, after correcting for the dust extinction, see \citealt{Gaiaext}) is about $4$~--~$6$ times larger in the metal-rich (${\rm [Fe/H]}>-1$) and/or Thin Disc-like subsample than in the metal-poor (${\rm [Fe/H]}>-1$) and/or Halo-like population. Although this simple analysis may be prone to observational biases, such as blends with spurious stars, artefacts, and reddening correction uncertainties, it suggests a potentially higher fraction of close binaries in the metal-rich RR Lyrae population. In any case, the population of RR Lyrae with unusual colours represents an interesting class of objects worth future follow-up studies. 

RR Lyrae with a white dwarf (WD) companion should not be uncommon either. In the Single Evolution Channel, given that the binary is sufficiently wide initially, the RR Lyrae progenitor may be the least massive of the two stars in the binary. Therefore, by the time the secondary star becomes RR Lyrae, the primary may have completed its evolution and become a WD. As follows from our population modelling, approximately $15\,\%$ of single-made binary RR Lyrae have a binary WD companion. In comparison, young and metal-rich binary-made RR Lyrae need to have relatively massive WD companions $M_{\rm WD}\gtrsim 0.8\,{\rm M}_\odot$ in order to form through stable mass transfer (lower-mass WD companions will typically lead to a common envelope phase). Therefore, only approximately $1\,\%$ of the binary-made RR Lyrae may have a WD companion.

\subsection{Comparison with Previous Results}   \label{sec:comparison}

\begin{figure}
    \centering
    \includegraphics[width=\columnwidth]{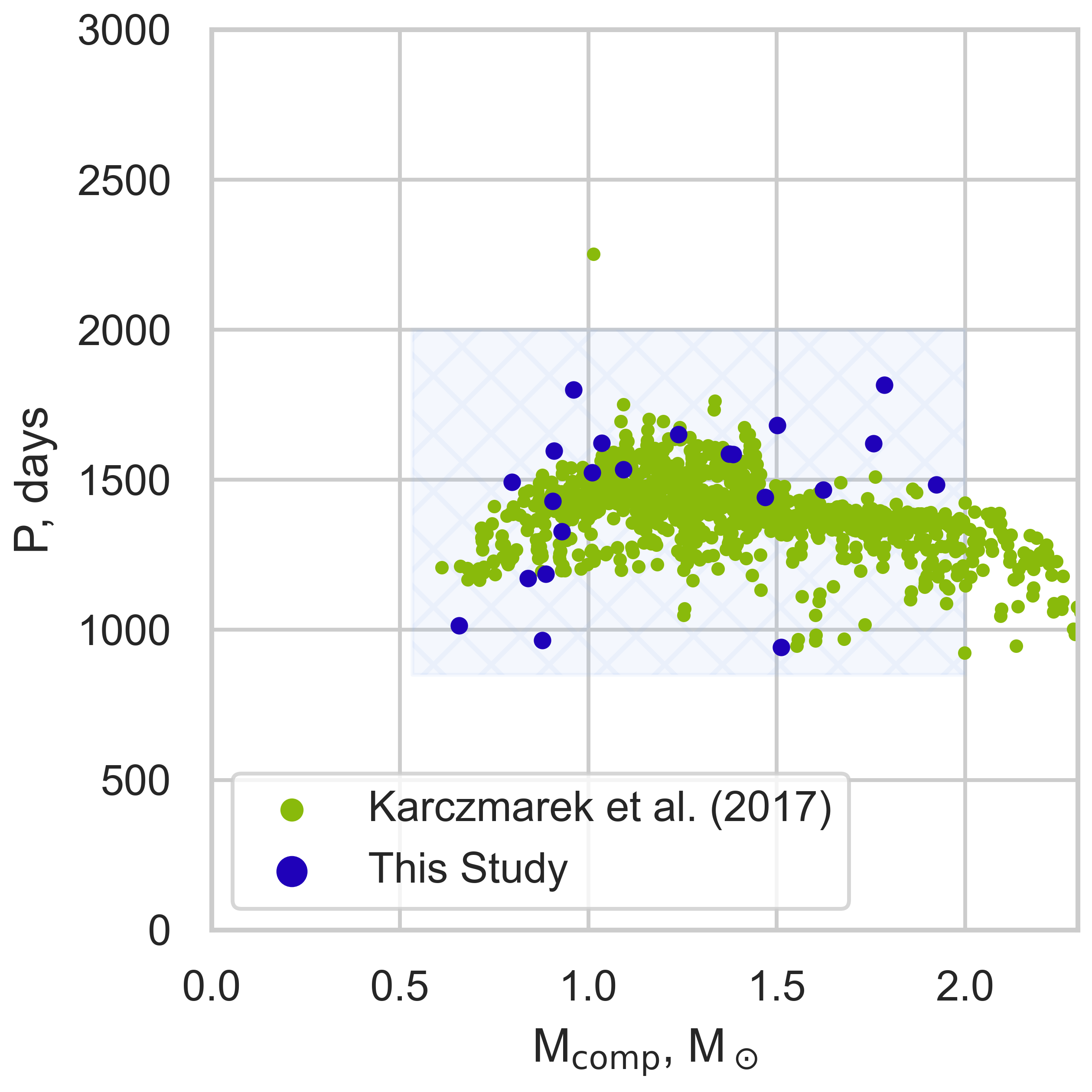}
    \caption{
    The orbital period and companion masses for the populations of binary-made metal-rich RR Lyrae from the \citet{BEPoccurrence} (green circles) study, overplotted by the systems from our study (blue circles).
    }
    \label{fig:MCompPeriodBEP}
\end{figure}

Our study builds on the work of \citet{BEPoccurrence}, which identified multiple possible channels of producing binary-made RR Lyrae pulsators. We focused on the systems formed through helium flash since they are structurally closest to the classical RR Lyrae and have similar core masses and surface temperatures. 

The main novelties and differences relative to the study of \citet{BEPoccurrence} are the use of the detailed stellar evolution code MESA, as opposed to the population synthesis code STARTRACK \citep{STARTRACK} used in \citet{BEPoccurrence}, and using a more realistic model for Galactic stellar populations. Both MESA and STARTRACK operate under a set of working assumptions, such as the choice of stellar wind model. However, MESA provides undeniable advantages in modelling binary mass transfer for low-mass stars. Firstly, it can self-consistently model the response of the star to mass loss and accretion, and secondly, MESA can follow stellar evolution through the He-flash. For these reasons, MESA is particularly well suited for studying the population of binary-made RR Lyrae stars.

Unlike \citet{BEPoccurrence}, we find that such systems constitute a significant fraction of all Galactic RR Lyrae and can fully explain the population of young and metal-rich RR Lyrae. In particular, while \citet{BEPoccurrence} predicted that binary-made RR Lyrae comprise only $0.8\,\%$, our study finds this fraction to be $10.1\%$. Despite the differences in binary evolution modelling, the dominant source of the discrepancy in the rates between \citet{BEPoccurrence} and our study is the Galactic model used.

\citet{BEPoccurrence} estimated their rates, assuming that $20\,\%$ of stars between $0.8$ and $0.9\,{\rm M}_\odot$  produce RR Lyrae. This model is representative of a metal-poor Halo-like population but not of the more metal-rich environments such as the Thin Disc and the Bulge (see e.g. Figure~\ref{fig:HRPhotSSE}). If the authors accounted for the metal-poor nature of the Halo and the metal-rich Thin Disc and the Bulge population sizes, their predicted fraction of binary-made RR Lyrae would be larger by approximately a factor of $10^2$, increasing to $80\,\%$ of all RR Lyrae. Other differences include the minimum stellar mass ($0.5\,{\rm M}_\odot$ in their work, $0.08\,{\rm M}_\odot$ in our model), the fraction of total and close binaries (they assume $100\,\%$ binary fraction, we used $25\,\%$ for the close binaries, and $45\,\%$ for all binary systems, see e.g. \citealt{Moe2019}). Moreover, they assumed that the RR Lyrae phase duration is $10\,{\rm Myr}$, whereas, according to our models, RR Lyrae live from $55.3\,{\rm Myr}$ for metal-poor stars to $92.3\,{\rm Myr}$ for metal-rich stars. Analysing the dataset provided by \citet{BEPoccurrence} and accounting for these differences, we estimate a production rate of $R_{\rm RRL}=0.45\,{\rm kyr}^{-1}$ for the binary-made RR Lyrae formed through the channel described in their paper. This value is accurate to within $30\,\%$ relative to the formation rate given in our Table~\ref{tab:Rates}.

Interestingly, despite the differences in binary modelling, the population properties in \citet{BEPoccurrence} generally agree with ours. In Figure~\ref{fig:MCompPeriodBEP}, we show the companion masses versus orbital periods for our dataset of binary-made RR Lyrae and the binaries in the \citet{BEPoccurrence} dataset producing RR Lyrae degenerately (with primary masses below $1.8\,{\rm M}_\odot$). The binary properties in both studies agree well. The discrepancy in the periods by about $20\,\%$ must be at least in part due to the alternative angular momentum prescription used in \citet{BEPoccurrence}. Furthermore, the Startrack code in general \citep{STARTRACK} assumes $50\,\%$ non-conservative mass transfer. In contrast, the observations of long-periods sdB binaries constrain mass transfer to be dominantly non-conservative \citep{Vos2020}. As a result, the companions in the \citet{BEPoccurrence} grow by $20$~--~$50\,\%$ compared to their initial masses, thereby extending the companion mass range above $2\,{\rm M}_\odot$ and expectedly affecting the final orbital periods. The use of solar metallicity for the whole population in \citet{BEPoccurrence} should have shifted portions of the period distribution by up to $50\,\%$ \citep{Vos2020}. In addition, due to non-calibrated helium ignition conditions and approximate estimates of the luminosity, the binary-made RR Lyrae in \citet{BEPoccurrence} are about $13$ per cent more massive than in our study, with some further implications for the orbit. For comparison, the range of periods of sdB binaries in \citet{Chen2013}, produced with the BSE code \citep{Hurley02}, underestimate the true orbital period range of the observed systems by up to a factor of several. Therefore, the relatively good agreement between the orbital properties in the \citet{BEPoccurrence} with our study is most likely related to some of the above effects cancelling each other out.

Given that \citet{BEPoccurrence} did not focus in-depth on the main channel of this study, it is noteworthy that there is a very good agreement in both the formation rates (under the correct treatment of the Galactic population) and the orbital properties of binary-made RR Lyrae. In our opinion, this provides additional support to both studies.

\subsection{Caveats}

The existence of binary metal-rich RR Lyrae in the Galactic disc is a robust prediction of binary stellar evolution. It is strongly supported by the fact that MESA-based simulations of interacting red giants, using the same model of mass transfer and the Galaxy as used here (as described in Section~\ref{sec:Method}), are in excellent agreement with the observed population of long-period subdwarf B stars. While binary companions of metal-rich RR Lyrae in the disc are still to be confirmed, below we review the possible caveats for our study.

\subsubsection{Stellar Winds} \label{sec:winds}

At very high metallicities, $\mathrm{[Fe/H]}\gtrsim-0.5$, low-mass core-helium-burning stars have too low surface temperatures, and they cluster on the so-called red clump. Postulating significant mass loss on the RGB is the only way to move these stars toward the instability strip. Indeed, \citet{Bono1997,Bono97b} postulated that young metal-rich RR Lyrae can be produced by progenitors with an initial mass within $1$~--~$2\,{\rm M}_\odot$ that have lost a considerable amount of mass during the ascent of the RGB ($M_{\rm lost} > 0.5\,{\rm M}_\odot$). 

In stellar evolution models, stellar winds represent a physical process responsible for mass loss along the RGB. This process is usually parameterised following \citet{reimers}
\begin{equation}
\dot{M}_\mathrm{RGB} = \eta \times 4 \times 10^{-13} \times \left( \frac{L}{\mathrm{L}_\odot} \right) 
 \left( \frac{M}{\mathrm{M}_\odot}  \right)
 \left( \frac{R}{\mathrm{R}_\odot} \right)^{-1} \, \mathrm{M}_\odot\, \mathrm{yr}^{-1},
 \label{eq:mdotrgb}
\end{equation}
where $L$, $M$ and $R$ are the bolometric stellar luminosity, the stellar mass and radius, and $\eta$ is the normalisation factor. The MIST tracks used in this work (Section \ref{sec:sse}) have been computed using $\eta=0.1$. Consequently, only a small amount of mass is lost along the RGB ($0.030$~--~$0.036\,\mathrm{M}_\odot$).
Typical $\eta$ values found in literature range between $0.1$ and $0.6$, while the amount of mass lost appears to increase with metallicity, but is constrained to be lower than $0.3\,{\rm M}_\odot$ \citep{Gratton10,Lei13,Miglio13,Origlia14,McDonald15,Savino18,Savino19,Savino20,Tailo20,Tailo21,Tailo22}.

To analyse the impact of mass loss on the formation of single-made RR Lyrae, we evolve a sample of stars using the public version of the population synthesis code \textssc{MOBSE}\footnote{\url{https://mobse-webpage.netlify.app}}\citep{Giacobbo18a,Giacobbo18b}. For each Galactic component (Table~\ref{tab:Besancon}), we have drawn $10^5$ stars sampling the initial masses, ages and metallicities as described in Section \ref{sec:mod}, and then evolved the stars up to the current time ($14\,{\rm Gyr}$). We estimate the  RR Lyrae formation efficiency, $f_\mathrm{RRL}$, as the percentage of core helium-burning stars currently in the IS (Equation~\ref{eq:is}), and we replicate this analysis for $6$ different values of $\eta$: $0.1$, $0.25$, $0.5$, $0.65$, $0.8$, and $1.0$.

For $\eta=0.1$, only old and metal-poor Halo stars can enter the instability strip ($f_\mathrm{RRL}\approx30\,\%$), confirming what we found with the MIST stellar tracks (Section \ref{sec:sse}). For larger values of the wind mass loss parameter ($\eta = 0.25$~--~$0.5$), the single-made RR Lyrae population is composed of the Halo and Thick Disc stars. At $\eta=0.25$, the ratio between the  halo ($f_\mathrm{RRL}\approx44\,\%$) and the disc ($f_\mathrm{RRL}\approx0.5\,\%$) formation efficiency is consistent with the one estimated by \cite{Layden1995b} in the solar neighbourhood  ($\approx 90\,\%$). The wind mass loss is significantly enhanced for more extreme values of $\eta (>0.5)$, and the stars lose from $0.3$ up to $0.48\,\mathrm{M}_\odot$ during the RGB. In this case, the RR Lyrae can also be produced in the Bulge and in the old portion of the Thin Disc ($>7\,{\rm Gyr}$) with $f_\mathrm{RRL}\lesssim 1.2\,\%$. At the same time, the Halo population is composed just of the few stars in the metal-rich tail ($f_\mathrm{RRL}\lesssim 2\,\%$). Therefore, no single value of $\eta$ can produce a realistic population of both metal-rich RR Lyrae in the disc and RR Lyrae in the Halo.

We further consider the possibility that the wind parameter $\eta$ depends on metallicity and increases sufficiently for higher-metallicity stars to produce a population of metal-rich RR Lyrae. Some caution is needed to interpret such results. First, the relation between $\eta$ and the effective amount of mass lost depends on the assumed stellar evolution tracks. \cite{Tailo20} found that high values of $\eta$ (up to $0.65$~--~$0.7$) are required to explain the horizontal branch of the most metal-rich globular clusters (GCs). However, the maximum mass lost in their model is $0.3\,\mathrm{M}_\odot$. This value is consistent with $\eta<0.6$ in the MOBSE stellar models. Assuming  $\eta=0.65$, the  RR Lyrae formation efficiency of the Thick Disc component  ($f_\mathrm{RRL}\approx21\%$) is comparable to that of the stellar Halo for $\eta<0.5$ 
($f_\mathrm{RRL}\approx20-40\%$). Given the significant mass difference between the two components ($M_\mathrm{Thick Disc}/M_\mathrm{Halo}>15$, see Table~\ref{tab:Besancon}), a large value of $\eta$ for the Thick Disc metallicity implies that the observed RR Lyrae stars should be mostly in the Thick Disc, in contradiction with the observations. Finally, no truly young population of RR Lyrae ($<7$ Gyr) can be produced through the Single Evolution Channel. Therefore, we can conclude that independently of the RGB wind-mass loss uncertainties, the Single Evolution Channel alone cannot explain the observed populations of young and metal-rich RR Lyrae stars.

Assuming significant helium enrichment, the amount of mass loss required to form such young and metal-rich RR Lyrae stars can be reduced \citep{Marsakov2019}. Helium-enriched stars are likely observed in the so-called second-generation populations in globular clusters  \citep{Caloi07,Tailo20,Dondoglio21} and the Bulge \citep{Minniti95}. Yet, there is no evidence of such stars in the Milky Way discs \citep{Reddy06,Karakas14} or He-enriched RR Lyrae stars in the Bulge \citep{Marconi18}, but see \citet{Savino20} for a different conclusion.

\subsubsection{Pulsation Properties} \label{sec:puls}

It is possible to retrieve a rough estimate of the fundamental mode pulsation periods based on the mean density formula, $P_{\rm pulsation} = Q \sqrt{\rho_\odot/\langle \rho \rangle}$, where the pulsation constant $Q = 0.04\,{\rm d}$ is representative of the characteristic range inferred from the observed pulsation periods, e.g., \citet{Stellingwerf1975,Jorgensen1967}.

We found that the binary-made RR Lyrae have systematically shorter pulsation periods ($0.15$~--~$0.3\,{\rm d}$) than single-made RR Lyrae ($0.35$~--~$0.45\,{\rm d}$). This difference is related to the fact that binary-made RR Lyrae are more compact than single-made RR Lyrae. While both single- and binary-made RR Lyrae undergo a He flash at nearly identical core mass distributions, their hydrogen envelope masses and metallicities differ. In particular, single-made RR Lyrae typically have cores of $0.495$~--~$0.515\,{\rm M}_\odot$ and envelopes of $0.22$~--~$0.24\,{\rm M}_\odot$. In comparison, binary-made RR Lyrae have cores of $0.460$~--~$0.475\,{\rm M}_\odot$ and envelopes of $0.04$~--~$0.07\,{\rm M}_\odot$.  Therefore, binary-made RR Lyrae always have relatively thin metal-rich envelopes, while single-made metal-poor RR Lyrae have thicker metal-poor envelopes. As a result, the latter are systematically less dense and, correspondingly, have longer pulsation periods.

The prevalence of short-period variables in the observed subsample of metal-rich RR Lyrae seems to match the predictions of our model  \citep[see, e.g.,][]{Muraveva18,Garofalo2022,Mullen22}. However, a comparison with the observed distributions of the RR Lyrae pulsation periods will require a much more detailed asteroseismic modelling than applying the mean density formulae \citep{Plachy2021}. In particular, such modelling could elucidate the effects of the systematic differences in masses, compositions, and distinct formation processes of binary-made and single-made RR Lyrae. Furthermore, additional studies are necessary to model the population and properties of the pulsators that ignited helium non-degenerately, which are not included in this study (see, e.g., \citealt{Smolec13,Chadid17}). In conclusion, this calculation tentatively suggests that shorter-period pulsators may be more promising candidates for binary-made RR Lyrae and, hence, for finding short-period companions. Conversely, single-made RR Lyrae, with their longer pulsation periods, may accommodate a wide range of binary companions.

Lastly, the answer to the general question of how a stellar pulsator emerges from the preceding binary mass transfer episode, and whether such pulsations may interfere with mass transfer or perhaps start after the end of mass transfer, remains unknown and is worth studying. MESA runs used in this study do not provide direct pulsation data, although these questions may be analysed, for example, with MESA and the Gyre code \citep{Townsend2013}. It is also worth examining the possible effects related to the change of initial helium abundance with metallicity, which is ignored in this study. The effects of non-solar helium abundances should be particularly pronounced for the Thick Disk and Bulge populations.

\subsubsection{Binary Orbital Periods}

Another possible caveat of our model is the final orbital periods of binary-made RR Lyrae. It is known that orbital periods resulting from binary mass transfer are sensitive to mass and angular momentum loss \citep{Soberman1997}. While the observed population of long-period sdB stars strongly constrains mass loss to be significantly non-conservative, the final orbital periods of sdB stars have no sensitivity to the angular momentum loss \citep{Chen2013,Vos2020}. Therefore, if the formation of binary-made RR Lyrae is sensitive to angular momentum loss, their orbital periods may change. If this is the case, measuring the orbital periods of binary-made RR Lyrae may put novel constraints on the angular momentum loss in binary evolution. Finally, independently of the angular momentum loss model, the population of binary-made RR Lyrae is robustly produced since these stars lie within the well-constrained continuous temperature distribution of binary stripped stars.

\subsubsection{Observed RR Lyrae Stars in Binary Systems}

The final caveat is that the two currently confirmed binary RR Lyrae come from channels different from the Binary Evolution Channel, which was the main focus of this study. In particular, the OGLE-BLG-RR Lyrae YR-02792 object has a mass of $0.26\,{\rm M}_\odot$ and cannot be produced by any considered channels. The detection of such an object is likely enabled by its short $15\,{\rm d}$ orbital period, to which binary microlensing detections are biased. Therefore, OGLE-BLG-RR Lyrae YR-02792 may represent a rarer but more easily detectable population from one of the channels described in \citet{BEPoccurrence}, e.g. RR Lyrae from the more massive progenitors than considered here ($M_{\rm progenitor}\gtrsim 2.1\,{\rm M}_\odot$) that ignited helium non-degenerately. Given the successful discovery of the OGLE-BLG-RR Lyrae YR-02792 object, the OGLE survey may be particularly well suited for detecting binary- and single-made RR Lyrae.

The other confirmed binary RR Lyrae TU UMa has Halo kinematics and, given its $\approx 8000\,{\rm d}$ orbital period, is certainly a single-made binary RR Lyrae from the Single Evolution Channel. Indeed, as detailed in Section~\ref{sec:Rates}, all the binary RR Lyrae in the Halo and $73\,\%$ of all the binary RR Lyrae in the Galaxy come from the Single Evolution Channel. Therefore, the only confirmed long-orbital-period binary RR Lyrae is also expected to come from this channel. Therefore, characterising binary-made RR Lyrae should be effectively calibrated by observationally characterising classical single-made RR Lyrae with a binary companion.

As follows from Section~\ref{sec:Result}, any binary metal-rich RR Lyrae should have companions with orbital periods of $1000$~--~$2000\,{\rm d}$. An encouraging, qualitative agreement may be seen upon inspecting the candidate systems from the \citet{Hajdu21} sample, as we show in Appendix~\ref{sec:BinRRLCandidates}. Generally, any further observational limits on the formation rates and orbital period distribution of long-period binary RR Lyrae will be highly constraining for the models of single-made and binary-made systems.

\subsection{Implications}

\label{sec:DiscussionImplications}

\subsubsection{RR Lyrae as Galactic tracers}

The main implication of this study is that metal-rich RR Lyrae in the Thin Disc and Bulge represent a young ($<7\,{\rm Gyr}$) stellar population. In particular, metal-rich RR Lyrae, the dominant RR Lyrae type in the solar neighbourhood, cannot serve as age indicators or tracers of old stellar populations. Conversely, the presence or absence of metal-rich RR Lyrae in a given sample may inform the Galactic population studies.

RR Lyrae as distance indicators should also be used with caution. While the metal-rich RR Lyrae are consistent with the (pulsation period)-luminosity-metallicity relation, their binary origins call for a more detailed study of their pulsations. In particular, the pulsation lightcurves may differ from those predicted by wind mass loss models for the classical single-made counterparts. Therefore, if binary-made RR Lyrae are misidentified as single-made RR Lyrae, and the pulsation models of single-made RR Lyrae are applied, the resulting pulsation periods may systematically differ. Furthermore, since metal-rich RR Lyrae have a notably lower luminosity than metal-poor RR Lyrae, they may affect the integrated mean metallicity estimates and increase the metallicity variance. In contrast, the local samples used to calibrate the pulsation period-luminosity-metallicity relation are better resolved and have a smaller metallicity variance. Therefore, the distance estimates might be biased and less accurate than expected. This is especially relevant when one considers regions of mixed metallicity in the same field, such as the Disc and Halo regions of other galaxies.

\subsubsection{RR Lyrae as probes of the binary evolution}

With their future characterisation, binary-made RR Lyrae will join the long- and short-period sdB binaries as the primary probes of red giant mass transfer. Each of the three populations may be modelled in detail, all within the same theoretical framework. Along with the mass distribution of short-period sdB binaries, observed orbital periods of binary-made RR Lyrae can potentially put novel constraints on the angular momentum loss in RG-MS binaries. The absence of $\lesssim 1\,{\rm d}$ orbital period RR Lyrae may be used to constrain common envelope evolution.

The existence of binary-made RR Lyrae implies that there are several related populations. In particular, the binaries with more massive A- and B-type accretors may lead to Be stars in binaries with RR Lyrae similar to the observed Be-sdO/B stars, e.g., \citet{Wang2021}. Furthermore, given that a non-negligible fraction of field binaries are members of triple systems, e.g., \citet{Toonen2016}, a subset of binary RR Lyrae will have a triple stellar companion. This configuration may potentially provide unique prospects for characterising binary RR Lyrae through timing. Moreover, as we showed in Section~\ref{sec:Result}, binary RR Lyrae may be produced from progenitors with initial mass ratios close to unity. Consequently, some RR Lyrae will have an evolved binary companion, and more rarely, another RR Lyrae in the same binary. Conversely, the absence of such systems may constrain the evolution of nearly equal mass binaries. For the same reason, a small fraction of binary RR Lyrae may be found with exotic companions, such as CO/ONe white dwarfs or neutron stars. Finally, similar mass transfer processes are relevant for forming the more massive Cepheid-like variables with possible implications for cosmology, e.g. see \citet{Karczmarek2022}.

It is also interesting to discuss the subsequent evolution of the binary single-made and binary-made RR Lyrae. In most binary-made RR Lyrae, the next evolutionary stage will happen when the RR Lyrae finishes core-helium burning and initiates shell-helium burning, by this time leaving the IS and appearing as an ordinary HB star. Having only small envelopes, the HB star will experience moderate expansion and eventually produce a \citep[potentially hybrid,][]{Zenati2019} white dwarf. Since most companions of binary-made RR Lyrae are stars above $0.8\,{\rm M}_\odot$ and at about $2$~--~$3\,{\rm AU}$ separations, a second phase of interaction, e.g., through a symbiotic binary phase is expected. In a subset of binary-made RR Lyrae with an evolved companion, an exotic phase may happen where the companion overflows the Roche lobe while the primary is still in the RR Lyrae phase. The accretion of material onto the RR Lyrae star and the resulting over-spinning may quench the RR Lyrae pulsations. 

In the systems resulting from single-made binary RR Lyrae, the companion usually has too little mass to evolve (apart from the previously mentioned exotic systems where the companion is an evolved star or is another RR Lyrae). However, the RR Lyrae descendant may, in less than $10\,\%$ of cases, overflow (or nearly overflow) its Roche lobe as an AGB star, interestingly enriching the metal-poor companions with CNO products.

\section{Conclusion}

In this work, we use detailed binary evolution simulations with the MESA code to analyse the plausibility of a new channel for RR Lyrae formation through envelope stripping in low-mass binary systems. We demonstrate that binary-made RR Lyrae can indeed be produced by partial mass stripping during the RGB phase. Our results not only confirm but also extend the findings of \citet{BEPoccurrence}, which are based on a different population synthesis code. Therefore, the existence of such RR Lyrae stars is a strong prediction of current models of binary stellar evolution.

Using a realistic model for Galactic stellar populations, we find that the binary-made RR Lyrae trace a population of metal-rich (${\rm [Fe/H]} \gtrsim -1$) and relatively young ($1$~--~$9\,{\rm Gyr}$) stars in the Thin Disc and the Bulge. These binary-made RR Lyrae could represent $\approx 10\,\%$ of the Galactic RR Lyrae population and up to $100\,\%$ of the observed metal-rich RR Lyrae (${\rm [Fe/H]}\gtrsim -1$). Considering the blend of various Galactic components, we estimate that RR Lyrae surveys targeting the bulge area could contain $\lesssim 10\%$ of binary-made RR Lyrae. In the region close to the Galactic plane, we expect a higher fraction ranging from $20$~--~$25\,\%$ within $3\,{\rm kpc}$ from the Galactic disc, and up to $70\,\%$ in the solar neighbourhood. The chemo-age-kinematic properties of the binary-made RR Lyrae in our simulations are consistent with the population of metal-rich disc-like RR Lyrae found in the Galactic Disc (see, e.g., \citealt{Zinn20,Prudil20,IB21}). We also predict that $38\,\%$ of the metal-poor RR Lyrae in the Stellar Halo are also in binaries but spanning a much wider range of orbital periods and not interacting or exchanging mass.

The Binary Evolution Channel is currently the only model that is able to explain both the formation and the properties of the observed metal-rich RR Lyrae in the Galactic disc comprehensively. The properties of such a population pose a significant challenge within the classical framework of RR Lyrae formation through single stellar evolution \citep{Bono1997,Catelan2004}.

Detecting companions to binary RR Lyrae provides crucial validation for the model introduced in this work. Currently, only two RR Lyrae have been confirmed to be in binary systems \citep{BEP,Liska16a}, but there exist several RR Lyrae binary candidates \citep{Kervella2019a,Kervella2019b,Hajdu21}. We did not find any significant discrepancies between the predictions of our model and the properties of the RR Lyrae binary candidates, which may be seen, for example, from the analysis done in Section~\ref{sec:RRLBinaryCats} and Appendix~\ref{sec:BinRRLCandidates}. Instead, we find encouraging agreement between observations and our model, such as the higher fraction of binary candidates within the metal-rich RR Lyrae population. At present, a detailed comparison between models and observations is limited due to the small-number statistics of both the data and the simulations. In future work, we aim to increase the number of simulations, expanding the mock RRL sample and exploring a wider parameter space.

We expect future observations (e.g. \gaia DR4) to reveal the companions of binary-made RR Lyrae ($P_\mathrm{orb}\lesssim 2000\,{\rm d}$). In the hypothetical case of conclusively ruling out the presence of companions of metal-rich RR Lyrae, observations of RR Lyrae stars would be in strong tension with the current models of binary stellar evolution. It would require a fundamental rethinking of many established aspects of both single and binary stellar evolution, such as mass transfer, mass loss, or evolution through the He flash. Since we actually observe a certain fraction of close binaries in the RR Lyrae progenitor populations \citep{Moe2019}, any model would have to provide a reason why binary-made RR Lyrae are not produced, whereas long-period sdB stars and stripped HB stars are produced. The rejection of the Binary Evolution Channel as an explanation of the observed metal-rich RR Lyrae would leave this enigmatic population without a satisfactory formation channel.

On the other hand, the confirmation of the formation channel presented in this work could signify a fundamental paradigm shift for RR Lyrae studies. RR Lyrae stars could be used to trace intermediate-age populations in the Galactic disc, in addition to tracing the Milky Way's oldest population, as has been done previously. Moreover, they could be included in the population of stars with stripped and partially stripped envelopes, representing a valuable class of objects against which to benchmark the theory of mass transfer in binary stars. Both the metal-poor and metal-rich populations of RR Lyrae stars will be explained within a unified framework, encompassing both single and binary stellar evolution.

\section*{Acknowledgements}

G.I. and V.B. are grateful to C. Tout for the illuminating discussion that kicked off this project. 
G.I. thanks Guglielmo Costa for the interesting and stimulating discussions 
and for the inclusion of the MIST tracks in SEVN. The authors thank the anonymous referee for their insightful comments.
A.B. acknowledges support for this project from the
European Union’s Horizon 2020 research and innovation
program under grant agreement No. 865932-ERC-SNeX.  The MESA simulations were performed on the resources provided by the Swedish National Infrastructure for Computing (SNIC) at the Lunarc cluster.
J.V. acknowledges support from the Grant Agency of the Czech Republic (GA\v{C}R 22-34467S). The Astronomical Institute Ond\v{r}ejov is supported by the project RVO:67985815.
G.I. acknowledges financial support from the European Research Council for the ERC Consolidator grant DEMOBLACK, under contract no. 770017.
G.I. and V.B. acknowledge support for this project by the Newton International Fellowship Alumni follow-on funding (2020, AL201003). G.I. thanks the Simons Foundation for financial support during his visit at the CCA where part of this work was performed. M.V. acknowledges support from the Fondecyt Regular No. 1211941 grant, funded by ANID.


\section*{Data availability}

The data underlying this article will be shared on reasonable request to the corresponding authors. A digital version of the tables in the Appendix can be found in the Zenodo repository 
\href{https://doi.org/10.5281/zenodo.10185011}
{10.5281/zenodo.10185011}.

\bibliographystyle{mnras}
\bibliography{main}

\appendix

\section{Comparison with the binary RR Lyrae candidates in the OGLE bulge field}
\label{sec:BinRRLCandidates}

\begin{figure}
    \centering
    \includegraphics[width=\columnwidth]{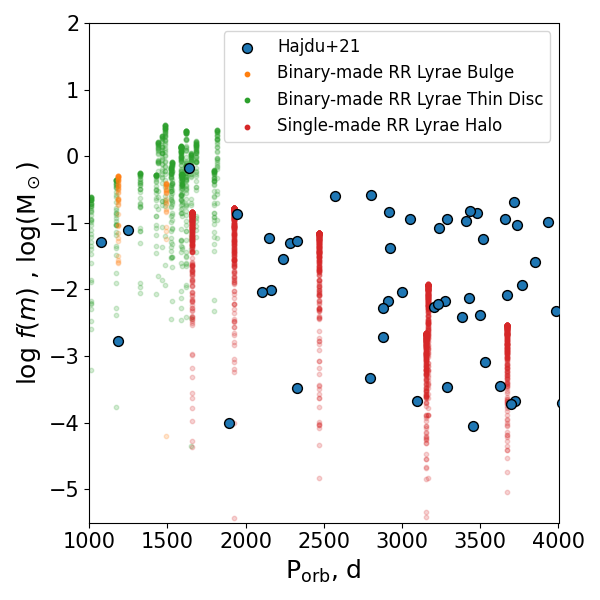}
    \caption{Orbital period versus logarithmic binary mass function (Equation~\ref{eq:massfunction}) for the binary candidates reported in \protect\cite{Hajdu21} and the binary RR Lyrae in our simulations. For each simulated binary RR Lyrae with an orbital period shorter than $4000\,{\rm d}$, we generate $1000$ realisations, randomly varying the inclination of the binary orbital plane, assuming an isotropic distribution for the inclination angle. The plot is limited to $4000\,{\rm d}$ (see text for details).}
    \label{fig:compareH21}
\end{figure}

The catalogue by \citet{Hajdu21} currently represents the largest homogeneous collection of binary RR Lyrae candidates with estimates of orbital period and mass function: 
\begin{equation}
f(m) = \frac{M^3_\mathrm{comp} \sin^3 i}{M^2_\mathrm{binary}},
\label{eq:massfunction}
\end{equation}
where $M_\mathrm{comp}$ is the mass of the companion of the RR Lyrae, $M_\mathrm{binary}$ is the total mass of the binary, and $\sin i$ is the sine of the inclination angle of the orbital plane with respect to the observer.

The sample analysed in \citet{Hajdu21}  (OGLE BULGE survey, \citealt{Soszynski2014}) contains only a few matches with the \gaia DR2 RR Lyrae analysed in \citet{IB21}, shown in Table~\ref{tab:ObsRates}. Yet, it is interesting to compare the observed properties of the binary candidates with the result of our simulations.

Seven per cent of the candidates in \citet{Hajdu21} have orbital periods consistent with the population of binary-made RR Lyrae in our simulations ($1000$~--~$2000\,{\rm d}$).  In the Bulge area explored by \citet{Hajdu21} (see figure 2 in \citealt{Soszynski2014}), our model predicts that among the $\approx 7\,\%$ of these binary RR Lyrae candidates there are binary-made RR Lyrae formed in the Bulge ($\approx 6\%$) and in the Thin Disc component ($\approx 1\%$). The rest of the candidate systems in \citet{Hajdu21} are long-period binaries ($\gtrsim 2000\,{\rm d}$) hosting single-made RR Lyrae from the Halo. However, the majority of the  Halo binary RR Lyrae in our simulations have very long periods ($\gtrsim 10^4\,{\rm d}$), while the longest estimated period in the \cite{Hajdu21} catalogue is $\approx 9000\,{\rm d}$.  This difference is likely due to the impossibility of detecting very long-period binaries using the method implemented in \citet{Hajdu21}. Limiting the comparison to orbital periods shorter than $4000\,{\rm d}$ (see Figure~\ref{fig:compareH21}), the fraction of the binary candidates in \citet{Hajdu21} with orbital periods shorter than $2000\,{\rm d}$ ($12\,\%$) is still comparable with the fraction of binary-made RR Lyrae in our simulations ($\approx 25\,\%$).

Since the mass function (Equation~\ref{eq:massfunction}) depends on the inclination angle of the binary plane with respect to the observer, we generate $1000$ realisations for each binary RR Lyrae in our simulation, randomly drawing the inclination angle assuming an isotropic distribution ($pdf(i) \propto \sin i$). Figure~\ref{fig:compareH21} shows the relation between the orbital period and the mass function for both the binary candidates from \cite{Hajdu21} and the random realisations for the binary in our simulations with periods shorter than $4000\,{\rm d}$. The binary-made RR Lyrae in our simulation have periods shorter than $2000\,{\rm d}$ and relatively high values of the mass function. This is due to the fact that the binary-made RR Lyrae tend to have more massive companions (see Section~\ref{sec:binprop}) and hence a higher upper limit of the mass function, i.e. when $i=90^\circ$. Similarly, the decreasing trend of $f(m)$  as a function of $P_\mathrm{orb}$ for the Halo binary RR Lyrae is due to the existence of an inverse correlation between the period and the mass of the companion.

Considering the complexity of the inner Galaxy (see, e.g., \citealt{Savino20,Queiroz21,Belokurov23}), the  $f(m)$-$P_\mathrm{orb}$ distribution of the binaries in our simulations shows an adequate qualitative agreement with the binary candidates from \cite{Hajdu21}. Interestingly, the binary candidate with the highest value of the mass function ($\log f(m)\approx0$) seems to be located in the area where we expect to find most of the binary-made RR Lyrae from the Thin Disc component. However, it seems that single-made RR Lyrae in our model show systematically lower values of the mass function than observed for the population of binary candidates with long periods. These candidate binaries host companion stars more massive than $0.6\,{\rm M}_\odot$ and could include RR Lyrae binaries with WD companions (see \citealt{Hajdu21}) that have not been investigated in this work. Alternatively, the discrepancy might relate either to observational bias favouring higher-mass companions or to the initial mass ratio distribution model used in this study.

\section{Table of binary-made RR Lyrae from simulations}

\begin{table*}
\begin{center}
\resizebox{\linewidth}{!}{%
\begin{tabular}{cccccccccccccc}
\hline
$\dfrac{\rm Age}{\rm Myr}$ & Galactic Bin & $\dfrac{M_{\rm proj}}{{\rm M}_\odot}$ &  $\dfrac{M_{\rm comp}}{{\rm M}_\odot}$ & $\dfrac{P_{\rm orb, init}}{\rm d}$ & [Fe/H] & $\dfrac{M_{\rm RRL}}{{\rm M}_\odot}$ & $\dfrac{M_{\rm comp, RRL}}{{\rm M}_\odot}$ & $\dfrac{P_{\rm orb, RRL}}{\rm d}$ & $\dfrac{M_{\rm core, RRL}}{{\rm M}_\odot}$ & $\dfrac{L_{\rm RRL}}{{\rm L}_\odot}$ & $\dfrac{T_{\rm eff}}{\rm K}$ & G & BP--RP \\
\hline
   1104 &          Thin Disc - Bin 3 & 1.855 & 1.758 &   132 &   0.216 &   0.478 &   1.758 &   1620 &                 0.452 &         15.4 &      8596 &  1.039 & 0.401 \\
   1160 &           Thin Disc - Bin 3  & 2.000 & 1.512 &    87 &  -0.028 &   0.469 &   1.512 &    941 &                 0.433 &         15.2 &      8058 &  1.330 & 0.252 \\
   1316 &           Thin Disc - Bin 3  & 1.643 & 1.091 &   372 &  -0.070 &   0.497 &   1.094 &   1533 &                 0.465 &         25.1 &      7555 &  1.099 & 0.311 \\
   1374 &           Thin Disc - Bin 3  & 1.803 & 1.784 &   143 &   0.006 &   0.488 &   1.925 &   1483 &                 0.460 &         18.4 &      7726 & -0.685 & 1.219 \\
   1452 &           Thin Disc - Bin 3  & 1.744 & 1.624 &   161 &  -0.071 &   0.493 &   1.624 &   1466 &                 0.463 &         21.3 &      7776 &  0.789 & 0.322 \\
   1456 &           Thin Disc - Bin 3  & 1.825 & 1.470 &   172 &   0.013 &   0.486 &   1.470 &   1441 &                 0.458 &         18.2 &      7553 &  1.195 & 0.343 \\
   1585 &           Thin Disc - Bin 3  & 1.628 & 1.239 &   298 &   0.073 &   0.492 &   1.239 &   1651 &                 0.463 &         22.2 &      7772 &  1.181 & 0.286 \\
   2157 &           Thin Disc - Bin 4 & 1.654 & 1.502 &   225 &   0.070 &   0.495 &   1.502 &   1681 &                 0.463 &         23.8 &      7537 &  0.901 & 0.372 \\
   2253 &           Thin Disc - Bin 4  & 1.593 & 1.563 &   256 &   0.088 &   0.492 &   1.787 &   1815 &                 0.463 &         20.7 &      7692 &  0.717 & 0.724 \\
   2430 &           Thin Disc - Bin 4  & 1.537 & 1.028 &   423 &   0.053 &   0.492 &   1.036 &   1621 &                 0.464 &         21.5 &      8490 &  1.353 & 0.139 \\
   2550 &           Thin Disc - Bin 4  & 1.498 & 1.375 &   255 &   0.010 &   0.493 &   1.375 &   1585 &                 0.465 &         25.2 &      8309 &  0.938 & 0.246 \\
   2734 &           Thin Disc - Bin 4  & 1.509 & 1.385 &   254 &   0.007 &   0.494 &   1.385 &   1585 &                 0.465 &         25.7 &      7620 &  0.870 & 0.344 \\
   3579 &           Thin Disc - Bin 5 & 1.278 & 0.829 &   481 &  -0.317 &   0.501 &   0.840 &   1171 &                 0.469 &         25.7 &      7641 &  1.119 & 0.275 \\
   4211 &           Thin Disc - Bin 5 & 1.287 & 0.893 &   547 &   0.043 &   0.494 &   0.909 &   1596 &                 0.464 &         23.0 &      7922 &  1.254 & 0.220 \\
   4282 &           Thin Disc - Bin 5 & 1.298 & 1.004 &   444 &  -0.031 &   0.497 &   1.010 &   1523 &                 0.466 &         24.4 &      7966 &  1.167 & 0.222 \\
   4743 &           Thin Disc - Bin 5 & 1.239 & 0.894 &   494 &  -0.098 &   0.494 &   0.906 &   1428 &                 0.467 &         28.8 &      8100 &  1.029 & 0.181 \\
   4934 &           Thin Disc - Bin 5 & 1.254 & 0.947 &   553 &   0.170 &   0.490 &   0.961 &   1800 &                 0.462 &         34.8 &      7410 &  0.779 & 0.329 \\
   8396 &           Thin Disc - Bin 7 & 0.948 & 0.877 &   430 &  -0.646 &   0.504 &   0.878 &    964 &                 0.474 &         27.7 &      7842 &  1.005 & 0.248 \\
   8455 &         Bulge & 1.053 & 0.780 &   660 &  -0.038 &   0.495 &   0.798 &   1492 &                 0.466 &         21.4 &      8237 &  1.374 & 0.155 \\
   8908 &           Thin Disc - Bin 7 & 1.019 & 0.925 &   503 &  -0.218 &   0.498 &   0.931 &   1328 &                 0.469 &         22.8 &      8135 &  1.241 & 0.201 \\
   9332 &           Thin Disc - Bin 7 & 0.948 & 0.640 &   616 &  -0.485 &   0.502 &   0.658 &   1013 &                 0.473 &         25.5 &      8132 &  1.182 & 0.164 \\
   9342 &         Bulge & 0.969 & 0.884 &   497 &  -0.360 &   0.500 &   0.888 &   1186 &                 0.471 &         23.1 &      8277 &  1.246 & 0.174 \\
   \hline
\end{tabular}}
\caption{The list of parameters for the binary-made RR Lyrae in our simulations. The columns show the ages of the systems since the zero-age main sequence, the Galactic bins from the Galactic model in Table~\ref{tab:Besancon}, the masses of the progenitors of the RR Lyrae and their companions, their orbital periods and metallicities, the masses of the RR Lyrae and the companions, and their orbital periods and RR Lyrae core masses, followed by the bolometric luminosities and effective temperatures of the RR Lyrae, along with their photometric Gaia G-bands and BP-RP colours. A digital version of this table can be found in the Zenodo repository 
\href{https://doi.org/10.5281/zenodo.10185011}
{10.5281/zenodo.10185011}.}
\label{tab:BMadeRRLDat}
\end{center}
\end{table*}

\begin{table*}
\begin{center}

\begin{tabular}{cccccccccc}
\hline
$\dfrac{\rm Age}{\rm Myr}$ & Galactic Bin & $\dfrac{M_{\rm proj}}{{\rm M}_\odot}$ & $\dfrac{M_{\rm RRL}}{{\rm M}_\odot}$ &  $\dfrac{M_{\rm RRL, comp}}{{\rm M}_\odot}$ & $\dfrac{P_{\rm orb, RRL}}{\rm d}$ & [Fe/H] & $\dfrac{M_{\rm core, RRL}}{{\rm M}_\odot}$ & $\dfrac{L_{\rm RRL}}{{\rm L}_\odot}$ & $\dfrac{T_{\rm eff}}{\rm K}$\\
\hline
14000 &          Halo &  0.763 & 0.728 &  0.069 &     589.4 & -1.820 & 0.497 &       48.8 &        7488 \\
     14000 &          Halo &  0.761 & 0.732 &  0.325 &     934.2 & -2.912 & 0.512 &       58.8 &        7679 \\
     14000 &          Halo &  0.764 & 0.732 &  0.014 &      1213 & -2.154 & 0.502 &       52.4 &        7698 \\
     14000 &          Halo &  0.763 & 0.729 &  0.691 &      1594 & -1.943 & 0.498 &       48.6 &        7866 \\
     14000 &          Halo &  0.766 & 0.733 &  0.597 &      1932 & -2.020 & 0.499 &       49.6 &        7859 \\
     14000 &          Halo &  0.766 & 0.732 &  0.013 &      3343 & -1.829 & 0.497 &       49.1 &        7443 \\
     14000 &          Halo &  0.762 & 0.727 &  0.549 &      4879 & -1.866 & 0.497 &       48.6 &        7669 \\
     14000 &          Halo &  0.758 & 0.725 &  0.226 &  2.81e+04 & -2.071 & 0.501 &       51.1 &        7727 \\
     14000 &          Halo &  0.757 & 0.729 &  0.340 & 4.373e+04 & -2.913 & 0.512 &       58.8 &        7706 \\
     14000 &          Halo &  0.765 & 0.733 &  0.188 & 4.436e+04 & -2.070 & 0.500 &       50.9 &        7776 \\
     14000 &          Halo &  0.762 & 0.730 &  0.339 & 5.097e+04 & -2.207 & 0.503 &       52.9 &        7706 \\
     14000 &          Halo &  0.768 & 0.734 &  0.473 & 6.043e+04 & -1.885 & 0.497 &       49.1 &        7604 \\
     14000 &          Halo &  0.762 & 0.731 &  0.688 & 6.241e+04 & -2.340 & 0.504 &       54.3 &        7678 \\
     14000 &          Halo &  0.768 & 0.737 &  0.410 & 7.536e+04 & -2.296 & 0.503 &       54.0 &        7657 \\
     14000 &          Halo &  0.758 & 0.728 &  0.071 & 1.178e+05 & -2.423 & 0.505 &       55.6 &        7589 \\
     14000 &          Halo &  0.770 & 0.736 &  0.540 & 1.162e+06 & -1.912 & 0.498 &       49.2 &        7640 \\
     14000 &          Halo &  0.762 & 0.733 &  0.576 & 1.948e+06 & -2.689 & 0.509 &       57.6 &        7645 \\
     14000 &          Halo &  0.758 & 0.724 &  0.680 & 4.254e+06 & -1.956 & 0.499 &       48.7 &        7886 \\
     14000 &          Halo &  0.760 & 0.729 &  0.510 & 1.173e+07 & -2.381 & 0.505 &       55.0 &        7623 \\
     14000 &          Halo &  0.758 & 0.728 &  0.615 & 3.958e+07 & -2.460 & 0.505 &       55.8 &        7597 \\
     14000 &          Halo &  0.756 & 0.722 &  0.365 &  5.94e+07 & -1.940 & 0.499 &       48.5 &        7901 \\
     14000 &          Halo &  0.761 & 0.726 &  0.387 & 6.804e+07 & -1.767 & 0.496 &       48.7 &        7338 \\
     14000 &          Halo &  0.764 & 0.731 &  0.138 & 1.053e+08 & -2.078 & 0.500 &       51.0 &        7765 \\
     14000 &          Halo &  0.759 & 0.728 &  0.437 & 1.221e+08 & -2.353 & 0.505 &       54.7 &        7635 \\
     14000 &          Halo &  0.760 & 0.726 &  0.285 & 1.628e+08 & -1.990 & 0.499 &       49.4 &        7838 \\
     14000 &          Halo &  0.757 & 0.722 &  0.056 & 4.455e+08 & -1.757 & 0.496 &       48.4 &        7369 \\
     14000 &          Halo &  0.763 & 0.729 &  0.256 & 8.769e+08 & -1.860 & 0.497 &       48.7 &        7626 \\
     14000 &          Halo &  0.755 & 0.724 &  0.374 & 9.064e+08 & -2.403 & 0.505 &       55.6 &        7592 \\
     14000 &          Halo &  0.761 & 0.725 &  0.357 & 1.016e+09 & -1.762 & 0.496 &       48.6 &        7321 \\
     14000 &          Halo &  0.764 & 0.729 &  0.598 & 1.931e+09 & -1.739 & 0.495 &       48.2 &        7181 \\
     14000 &          Halo &  0.757 & 0.724 &  0.408 & 2.321e+09 & -2.088 & 0.501 &       51.4 &        7725 \\
     14000 &          Halo &  0.767 & 0.733 &  0.404 & 2.567e+09 & -1.840 & 0.497 &       49.1 &        7467 \\
   \hline
\end{tabular}
\caption{The list of parameters for a representative subset of single-made binary RR Lyrae in our simulations. The columns show the ages of the systems since the zero-age main sequence, the Galactic bins from the Galactic model in Table~\ref{tab:Besancon}, the masses of the RR Lyrae progenitors, the present-day masses of the RR Lyrae and their companions, their orbital periods and metallicities, the RR Lyrae core masses, and the bolometric luminosities and effective temperatures of the RR Lyrae. A digital version of this table can be found in the Zenodo repository 
\href{https://doi.org/10.5281/zenodo.10185011}
{10.5281/zenodo.10185011}.}
\label{tab:SMadeRRLDat}
\end{center}
\end{table*}

In Table~\ref{tab:BMadeRRLDat}, we list the properties of binary-made RR Lyrae produced in our MESA simulations. For comparison, in Table~\ref{tab:SMadeRRLDat}, we show the properties for a subset of single-made binary RR Lyrae.

\bsp	
\label{lastpage}
\end{document}